\documentclass[showpacs,showkeys,12pt,preprint,preprintnumbers,nofootinbib,groupedaddress,superscriptaddress]{revtex4}
\usepackage{amsmath,amssymb,amsfonts,color}
\usepackage[dvipdfm]{graphicx}

\newcommand{\eq}[1]{Eq.~(\ref{#1})}
\newcommand{\fsm}{f}
\newcommand{\Fmc}{q} 
\newcommand{\ep}{\epsilon}
\newcommand{\bep}{\bar{\epsilon}}
\newcommand{\bea}{\begin{eqnarray}}
\newcommand{\eea}{\end{eqnarray}}
\newcommand{\oQ}{\overline{Q}}
\newcommand{\oL}{\overline{L}}
\newcommand{\qf}{{{\mathcal{Q}}_F}}

\newcommand{\ts}{\text{s}}
\newcommand{\tc}{\text{c}}
\newcommand{\ttg}{\text{t}}
\newcommand{\T}{\text{T}}
\newcommand{\Po}{\text{P}}

\newcommand{\ckm}{{V_{\small{\rm{CKM}}}}}

\newcommand{\nn}{\nonumber}
\begin{document}

\vspace*{1cm}
\title{Phenomenology of flavon fields at the LHC}
\author{Koji Tsumura}
\email{ktsumura@ictp.it} \affiliation{The Abdus Salam International Centre for Theoretical Physics, Strada Costiera 11, 34151 Trieste, Italy}
\author{Liliana Velasco-Sevilla$^1$}
\email{lvelasco@ictp.it} 
\preprint{IC/2009/078} \pacs{12.60Fr,12.15Ff} \keywords{Collider Phenomenology, quark and lepton masses and mixing, family symmetries}
%

\begin{abstract}
We study low energy constraints from flavour violating processes, production and decay at the LHC of a scalar field $\varphi$ ({\it flavon}) associated to the breaking of a non supersymmetric Abelian family symmetry at the TeV scale. This symmetry is constrained to reproduce fermion masses and mixing, up to $O(1)$ coefficients. The non-supersymmetric gauged $U(1)$ models considered are severely restricted by cancellation of anomalies and LEP bounds on contact interactions, consequently its phenomenology is out of the LHC reach. We therefore introduce an effective $U(1)$ which is not gauged and it is  broken explicitly by a CP odd term at the TeV scale. This helps us to explore flavour violating processes, production and decay at the LHC for these kind of light scalars. In this context we first study the constraints on the flavon mass and its vacuum expectation value from low energy flavour changing processes such as $\mu\to e\gamma$. We find that a flavon of about $m_\varphi \lesssim 150$ GeV could be experimentally allowed. These kind of flavons could be significantly generated at the LHC via the gluon fusion mechanism and the single top production channel $gu\to t\varphi$. The produced flavons can have characteristic decay modes such as $t \bar{c} $ for $m_\varphi \gtrsim m_t$, and $\tau \bar{\mu}$ for $m_\varphi \lesssim m_t$, which could be effectively useful to detect flavons.

\end{abstract}

\maketitle

\section{Introduction}

Family symmetries are introduced to explain the hierarchy of fermion masses and mixing in the standard model (SM) and usually the breaking scale of this symmetry is set up near the Grand Unification or the Planck scales, $M_\text{GUT}$ and $M_\text{P}$ respectively. In these scenarios, often the possible extra-particles required by these family symmetries decouple at those high energies and hence no observable consequences, except for an explanation of the values of the Yukawa couplings, appear at the electroweak scale (EW). Giving the exciting opportunity for particle discoveries at the LHC, we study the possibility of light scalars breaking family symmetries within few TeVs.
To this end, we present a non-supersymmetric effective approach where we start with a Lagrangian describing the interaction of the scalars breaking the hypothetical family symmetry, that henceforward we call {\it flavons}, at a scale $M_\text{F}$ that is set in the TeV range. The coupling of these scalars to SM fermions and the Higgs boson generates hierarchical Yukawa couplings {\it \`a la}  Froggatt-Nielsen (FN) \cite{Froggatt:1978nt} and controls the flavour violating processes induced.

The paper is organized as follows: in $\S$ II we study the construction of models based on family symmetries in this context, where in order to simplify the analysis we just consider $U(1)$ groups. From exhaustive studies on the construction of family symmetries with such groups at high scales \cite{ref:u1s}, in spite of its elegance, we know however their limited predictability. One of the most severe being the impossibility in establishing relations among the $O(1)$ coefficients associated to the effective Yukawa couplings produced by powers of the vacuum expectation value (vev) of the flavons. This problem could be somehow alleviated in non-Abelian models \cite{ref:non-Ab}. However, here we would like to probe the possible size of flavon production cross sections and branching ratios of its decays at the TeV scale and for this illustrative purpose it is enough to consider Abelian family symmetries.

 We find that for a gauged group $G_{\text{F}}=U(1)_{\text {F}}$ coupling to all the SM fermions or for $G_{\text{F}}=U(1)_{\text{F}_1}\times U(1)_{\text{F}_2}$ with some of the SM fermions coupling to $U(1)_{\text{F}_1}$ and others to $U(1)_{\text{F}_2}$, the masses of the scalars and vector bosons could be light, even below the TeV range. However the flavour violating processes, production and decay of this scenario are not relevant below the TeV scale and hence out of the scope at the LHC. Since we have worked in a non supersymmetric context, the cancellation of anomalies it is of a different nature to those supersymmetric models at high energies (e.g. \cite{ref:u1s}) and hence the solutions found for Yukawa couplings are different.

We then introduce an effective global $U(1)_{\text {F}}$ which is explicitly broken below 1 TeV, which avoids the appearance of a Nambu-Goldstone boson, but gives rise to a pseudo-Nambu-Goldstone boson, PNGB, which controls the flavour violating processes. In $\S$ III we present the constraints on the flavour violating parameters from a fit to fermion masses and mixing. In $\S$ IV we derive the branching ratios of the flavour violating processes mediated by the PNGB mentioned above. Its production and decay modes at the LHC are studied in $\S$ V. Finally in $\S$ VI we summarize our results.

%

\section{Theoretical motivations and constraints}
The FN mechanism introduces scalars charged under a family symmetry with group $G_\text{F}$ that can couple to the SM fermions, which are also charged under this symmetry. When this is broken we can obtain effective Yukawa couplings with a hierarchy controlled by the charges and the vev of the flavons breaking $G_\text{F}$. Let us start our discussion with the effective mass Lagrangian for SM fermions given in terms of flavons of an Abelian theory:
\bea
\label{eq:Lfn}
-{\mathcal L}_\text{FN} & =&
{\displaystyle\sum_n{\overline {\mathcal{F}}_{Li}  {\mathcal{F}}_{Rj}\Phi {c^{\mathcal{F}}_{\varphi^{\mathcal{F}}_{n}}}_{ij}  \left(\frac{
\varphi^{\mathcal{F}}_n}{\Lambda_{\varphi_n^{\mathcal{F}}}}\right)^{p^{\mathcal{F}}_{n ij}}  }} + {\text{H.c.}},
\eea
where there could be a different number of $n$ flavons $\varphi_n$ coupling to the different SM fermions $\mathcal{F}$. The coefficients ${c^{\mathcal{F}}_{\varphi^{\mathcal{F}}_{n}}}_{ij}$ are of $O(1)$ and in principle can be different for each coupling and kind of fermion.
The scales $\Lambda_{\varphi}$ could be associated with masses of extra fermions coupling to the SM ones and to the scalars $\varphi$, but such that they do not play an important role in the phenomenology below the TeV scale.

The predictability of a particular family symmetry consists in explaining all the fermion masses and mixing with less parameters than predictions. Therefore in practice predictive family symmetries have few of flavons and generically all of them couple to all fermions, for this reason we consider that only one flavon in each sector (i.e. one for quarks and other for leptons) controls the flavour changing neutral current (FCNC) processes below the TeV scale.

\subsection{Global  $U(1)$  symmetries \label{subsec:globalu1}}

Here we would like to consider an approximate global $U(1)$ symmetry with an explicit CP odd breaking term which gives mass to a PNGB, $a$, at the TeV scale. Collider bounds for the effective $GGa$ and $\gamma\gamma a$ couplings  have been looked for, assuming that the scalars $a$ couple only to photons, $\gamma$, and gluons, $G$ \cite{Kleban:2005rj}. This analysis expects symmetry breaking scales above 1 TeV.
 However, in our case, the scalar $\varphi$ also couples to fermions via the Lagrangian of \eq{eq:Lfn} and such couplings provide additional decay channels that where not considered in \cite{Kleban:2005rj}.
The goal of this work is to probe the flavon coupling scales through flavour changing violating processes, production and decay of flavons and we explore ranges from few hundreds of GeVs up to 1 TeV. Studies of this kind have been considered before in \cite{Dorsner:2002wi,Nandi:2009gr}, here we focus on more general possibilities and production and decay rates of such flavons. If flavons are lighter than the $Z$ boson, the decays $Z\to f \bar f \varphi$ at tree level and $Z\to \gamma \varphi$ at one-loop level could take place. However, the lower bound on the flavon mass is not robust and strongly depends on the particular models considered.

On the theoretical side, a symmetry breaking scale should correspond to a definitive mechanism or interaction. We do not have an answer to such in this paper, here we just would like to probe scales in the range $\sim (200,1000)$ GeV and, as stressed above, for it we use flavour changing violating processes, production and decay of flavons. Remember that if the interaction giving rise to the symmetry breaking of a global $U(1)$ were of a gravitational origin, this would severely constrain the mass term of the PNGBs, $m_a$, and the scale $f$ at which such a symmetry should be broken {\footnote{Just for the axion $m_a$ and $f_a$ are related but for a general scalar or pseudo-scalar there is no such relation.}} . In this case, such scale can be determined from its coupling to two photons or two gluon fields: $\mathcal{L}=\frac{1}{8} g_\gamma \epsilon_{\mu\nu\alpha\beta} F^{\mu \nu} F^{\alpha\beta} ~ a + \frac{1}{8} g_g \epsilon_{\mu\nu\alpha\beta} G^{\mu \nu} G^{\alpha\beta} ~ a$, where $g_\gamma= 8 \alpha/(\pi f_\gamma)$ and $g_g= 3 \alpha_s/(\pi f_g)$. A variety of experiments have explored scalar or pseudo-scalars with these properties but most of them are just sensitive to $m_a<1$ GeV \cite{Battesti:2007um}. The Super KEKB upgrade to KEKB is expected to improve an existing bound on $g_\gamma$ up to $<1.9 \times 10^{-6}$ GeV${}^{-1}$ for a mass $m_a<0.13$ GeV. This translates into a breaking scale of up to $f_\gamma >3 \times 10^4$ GeV. Therefore here we consider that these effects do not play a role in the phenomenology between SM fermions and the flavons.

In our effective approach we consider that the vacuum expectation value, naturally of the order of $f$, is obtained from the following scalar potential
\begin{align}
V(\Phi,\phi) =
-\mu^2\left|\Phi\right|^2+\lambda\left|\Phi\right|^4
-\mu_\varphi^2\left|\varphi\right|^2+\lambda_\varphi\left|\varphi\right|^4
+2\lambda'\left|\Phi\right|^2\left|\varphi\right|^2
-\frac{M^2}4(\varphi^2+{\varphi^*}^2),
\label{eq:scalar_u1psglob}
\end{align}
where $\Phi$ is the SM Higgs doublet and the term proportional to $M^2$ is the explicit $U(1)$ breaking parameter and the complex flavon can be parameterised as
\bea
\varphi=\frac{(v_{\varphi}+H_f+i A_f)}{\sqrt2}.
\eea
Here $H_f$ is a CP even flavon which may be integrated out, and $A_f$ is a CP odd flavon, the PNGB, whose mass is determined by the symmetry breaking parameter $M^2$, i.e., $m_{A_f}^2=M^2$. In general, we could consider the following mass matrix for the SM Higgs boson and the flavon
\begin{align}
{\mathcal{M}}^2=\frac{1}{2}\frac{\partial^2  V }{\partial \phi_i \partial \phi_j}|_{\phi_i=\frac{v_i}{\sqrt{2}}}=
\frac{\partial^2 \langle V \rangle}{\partial v_x \partial v_y} =
2 \begin{pmatrix}
\lambda v^2 & \lambda' v v_\varphi\\
\lambda' v v_\varphi & \ \ \lambda_\varphi v_\varphi^2
\end{pmatrix},
\end{align}
which induces a mixing between $\Phi$ and $\varphi$. Hence the effects of the CP even flavon could also be observed in Higgs boson phenomena even if the CP even flavon was heavy enough. A detailed analysis via the Higgs boson requires precise measurements of Higgs couplings, which is beyond the scope of this paper.  However we can see that even when not neglecting such a mixing there would not be worries due to the radiative corrections to the Higgs mass via the term $2\lambda' |\Phi|^2 |\varphi|^2$. This term gives a contribution $\Delta m^2_H\sim\frac{\lambda'}{16\pi^2}  m_\varphi^2$, which is a small contribution, as long as $\lambda'$ is small and $m_\varphi$ is kept below the TeV scale. Of course, quadratic divergences appear in this theory but are of the same type as the ones present in the SM.

We just then concentrate on the phenomenology of the flavon without such mixing. In such case the masses of the CP even and odd scalars are given by
\bea
m_{H_f}^2&=&2\lambda_\varphi v_\varphi^2 \nn\\
m_{A_f}^2&=&M^2
\eea
 We note that even if the vev of the flavon, $v_{\varphi}$, is heavy, its CP-odd component $A_f$, a PNGB, could be lighter than its CP-even part, and therefore the contribution to processes below $v_\varphi$ from $H_f$ would be subdominant. Other PNGBs have been introduced as scalars controlling quarks and mixing at TeV scale in the context of the little Higgs models \cite{littlehiggsflavons}. In our context, $M$ can be taken at the electroweak scale, taking $M_{A_f}>M_Z$ to avoid $Z$ decay constraints on flavon mass but we discuss $M_{A_f}$ up to 1 TeV {\footnote{We would like to emphasize that in theories where there is a spontaneous symmetry breaking involved in global symmetries through non SM singlets $\varphi$'s, there are very dangerous contributions to the invisible decay width of the $Z$ boson \cite{GonzalezGarcia:1989zh}, of course valid for $M_Z> M$. Clearly this is not a source of worries for this scenario.}}  
For the values of $M^2$ that we are considering, long range forces give a very weak constraint \cite{longrange} which is compatible with assuming $M$ above the electroweak scale.

 We remark that no cosmological defects appear from the potential of \eq{eq:scalar_u1psglob} because the explicit breaking term lifts up the degeneracy of vacua, and the resulting conditions for the minimisation of the potential leaves just one solution for $v_\varphi$.

Although the coupling strengths of flavons depend on the symmetry breaking scale, the motivation of producing the Yukawa couplings at low scale through the couplings of scalars whose signatures could be studied at collider experiments is exciting. Approximate continuous global symmetries could arise in scenarios beyond the standard model \cite{Burgess:2008ri}. However at the end, the ultimate purpose of this paper is to probe simple family symmetries, whose phenomenology is determined by a single flavon, through its flavon production and decay at colliders without stating the origin of family symmetry breaking.
%
%
\begin{table}
\begin{center}
\begin{tabular}{|l| l l l l l l|}
\hline
\multicolumn{7}{|c|}{$U(1)_F$ charges}\\
\hline
Field  & ${\overline{Q}_L}_i$ \ & ${d_R}_i$\ & ${u_R}_i$ \ &
${\overline{L}_L}_i$ \ & ${e_R}_i$\  & ${\nu_R}_i$\ \\
Charge \ & ${\Fmc_L^{\overline{Q}}}_i$ \ & ${\Fmc_R^d}_i$ \ & ${\Fmc_R^u}_i$\ &
${\Fmc_L^{\overline{L}}}_i$ \ & ${\Fmc_R^e}_i$\  & ${\Fmc_R^\nu}_i$\  \\
\hline
\end{tabular}
\end{center}
\label{tbl:U1_not_ch}
\caption{\small Notation for $U(1)_F$ charges for fermions.}
\end{table}
%
%

In the phenomenological analysis that we present in \S \ref{subs:MIA_1_flavon} we consider the effective mass Lagrangian as a result of a single flavon field, whose potential is described by \eq{eq:scalar_u1psglob}, and the coupling to fermions by
\begin{align}
-{\mathcal L}_\text{FN} & =
c_{ij}^\ell \overline{L}_i{e_R^{}}_j\Phi\left(\frac{\varphi}{\Lambda}\right)^{-\qf\left(\Fmc^{\oL}_{Li}+\Fmc_{Rj}^e\right)}
+ c_{ij}^d \overline{Q}_i{d_R^{}}_j\Phi\left(\frac {\varphi}{\Lambda}\right)^{-\qf\left(\Fmc_{Li}^{\oQ}+\Fmc_{Rj}^d\right)}\nonumber\\
&+ c_{ij}^u \overline{Q}_i{u_R^{}}_j\tilde\Phi\left(\frac{\varphi}{\Lambda}\right)^{-\qf\left( \Fmc_{Li}^{\oQ}+\Fmc_{Rj}^u \right) }
+\text{H.c.},
\label{eq:mass_lag}
\end{align}
where $\varphi$ and $v_\varphi$ are the flavon field and its vacuum expectation value, respectively. Finally ${\Fmc^f_{L/R}}_i$ are the $U(1)_F$ charges of the corresponding kind of fermion, following the notation of Table I,  and $\qf$ is the charge of $\varphi$, normalized to $\pm$ 1. We can choose $\qf=-1$ and hence the hierarchical Yukawa matrices are generated as a consequence of the breaking of the family symmetry and have the form
\begin{align}
\label{eq:yuk_cps_oneu}
Y^{\fsm}_{ij} =
c_{ij}^{\fsm} \left(\frac{v_\varphi}{\sqrt{2} \Lambda}\right)^{\left({\Fmc_L^{\overline{\fsm}}}_i+{\Fmc_R^\fsm}_j\right)},
\end{align}
where the coefficients $c^f_{ij}$ are of order one. In order to diagonalise the mass matrix, the electroweak fields are redefined as
\bea
&& L\to U_L^e L,\quad e_R^{}\to U_R^e e_R^{}\quad \Rightarrow  \quad Y^e={U_L^e}^\dag Y^e_\text{diag} U_R^e,\nn\\
&& Q\to U_L^q Q,\quad q_R^{}\to U_R^q q_R^{}\quad \Rightarrow  \quad Y^q={U_L^q}^\dag Y^q_\text{diag} U_R^q,
\eea
where $Y^e_\text{diag}=\text{diag}(m_e,m_\mu,m_\tau)\frac{\sqrt2}{v}$, analogously for the case of quarks.

We then obtain the flavour diagonal Yukawa interaction and the FCNC interaction with a flavon field,
\begin{align}
-{\mathcal L}^\text{eff}_{\varphi} & =
m^{\fsm}_{i}\,\overline{\fsm_L^{}}_i{\fsm_R^{}}_i\left(1+\frac{h}{v}\right)
+\kappa^\fsm_{ij}\,\overline{\fsm_L^{}}_i{\fsm_R^{}}_j
\left(\frac{\varphi}{v_\varphi/\sqrt{2}}\right)
+\text{H.c.},\label{Eq:L_eff}
\end{align}
where $h$ is the Higgs boson. Below the scale at which the CP even part of $\varphi$ decouples, there is a contribution to this effective Lagrangian from the five-dimensional operators involving the CP odd part, which does not acquire a vev. These however introduce just negligibly effects in the phenomenology below the decoupling scale, $\Lambda$, since it is suppressed by powers of $\varphi/\Lambda$. When only one flavon couples to each kind of fermion then the flavour violating matrices $\kappa^\fsm_{ij}$ can be decomposed into left- and right-handed parts as:
\begin{align}
\kappa^\fsm_{ij} & = \left[ m^\fsm
_{j}\sum_k{\Fmc_L^{\fsm}}_k
\left({U_L^\fsm}\right)_{ik}\left({U^\fsm_L}\right)^*_{jk}
+m^\fsm_{i}\sum_k{\Fmc_R^{\fsm}}_k
\left({U_R^\fsm}\right)_{ik}\left({U^\fsm_R}\right)^*_{jk}
\right].
\label{Eq:k_ij}
\end{align}
The separation of the flavour violating matrices in left and right components is one of the specific features of the global $U(1)_F$ with one flavon coupling to each kind of fermion. We will explore the consequences of this separation in the next sections. In Tables II and III we present two plausible sets of $U(1)_F$ charges that could reproduce the hierarchy of fermion masses and mixing. From the construction \cite{ex_susyu1} of flavour models in supersymmetric GUTs we know that it is very difficult to reproduce fermion masses and mixing with a single $U(1)$ group but we explore this possibility which it is enough to illustrate the size of flavour violation and flavon production in this context.
\subsection{Gauged $U(1)$  symmetries \label{sbsc:gaguedmod}}
%
%
On the theoretical side these symmetries are severely restricted by the cancellation of anomalies and on the experimental one, the  masses of the associated gauge bosons and its couplings to SM particles are sorely constrained by LEP and Tevatron searches. Hence these two aspects together can be used as a guideline for constraining and analysing the signatures of these models. In what it follows we first present the conditions from anomaly cancellations, which give different solutions to the supersymmetric cases, and then the restrictions from the couplings to the extra gauge bosons. Remember that we would like to generate the hierarchy of masses and mixing with different  $U(1)$ charges for fermions of different families, therefore the study and phenomenology of this differs from the ``universal'' $U(1)$ charges scenarios that have been widely studied \cite{Amsler:2008zzb}. In addition, since the models are not supersymmetric, the cancellation of anomalies has different solutions to the supersymmetric ones \cite{ref:u1s}.

\subsubsection{Conditions from cancellation of anomalies.}

Here we briefly describe the way anomaly-free $U(1)_F$ gauge symmetries, with at least one scalar that can be identified with a flavon near the TeV scale, could be constructed.
With the assumption that all the generations of a given fermion type will carry the same $U(1)$ charge, the only extra symmetries beyond the SM that its particles are allowed to have are $U(1)$ symmetries whose generators are linear combinations of the hypercharge and the difference $B-L$ \cite{Weinberg:1996kr} {\footnote {In particular it is well known the solution $U(1)_{B-L}$ which requires the inclusion of three extra fermions with zero hypercharge that can be identified with the right-handed neutrinos.}}.
 When we give to fermions of different families different charges, in this non supersymmetric context, it is not possible to achieve just with one gauged $U(1)$ a complete description of the hierarchies of masses and mixing.

The cancellation conditions of triangle mixed anomalies, with external gauge boson lines and internal lines of a SM fermion, of the type $U(1)_{\rm{F}}-G^i_{\rm{SM}}-G^i_{\rm{SM}}$, where $G^i_{\rm{SM}}=U(1)_{\rm{Y}}$, $SU(2)_{\rm{Y}}$, $SU(3)_{\rm{C}}$, are given by $A^i=\frac{1}{2}\text{Tr}\left[T^F_c \left\{T^i_a,T^i_c \right\}\right]=0$. Here $T^i_a$ are the generators of the SM groups and $T^F_a$ of $U(1)_{\rm{F}}$ and we have used the normalizations $\left\{Y,Y \right\}=2Y^2$ and $\left\{T_a,T_b \right\}=\delta_{ab}$. Also $A_F=\frac{1}{2} \left[T^F_c \left\{T^F_a,T^F_c \right\}\right]$ and $\text{Tr}[T_a^{U(1)_{F}}]$ must cancel. We rewrite the familiar anomaly cancellation expressions in terms of the family dependent charges:
\bea
6 A_1&=&\sum^{3}_{i=1}{\Fmc^{\oQ}_L}_i+ 8 {\Fmc_R}^u_i + 2 {\Fmc_R}_i^d + 3 {\Fmc^{\oL}_L}_i +6 {\Fmc_R}^e_i\nonumber\\
2 A_2&=&\sum^{3}_{i=1} 3{\Fmc^{\oQ}_L}_i+  {\Fmc^{\oL}_L}_i\nn\\
2 A_3&=&\sum^{3}_{i=1} 2{\Fmc^{\oQ}_L}_i+ {\Fmc_R}^u_i +  {\Fmc_R}_i^d\nn\\
2 A_F&=&\sum^{3}_{i=1} {\Fmc^{\oQ}_L}^2_i -2 {{\Fmc_R}^u_i}^2 +  {\Fmc_i^d}^2 - {\Fmc^{\oL}}_i^2 + {{\Fmc_R}^e_i}^2.
\label{eq:anom_us_a}
\eea
When the charge of the Higgs boson is zero, the parametric sums \cite{Jain:1994hd} that solve the equations above, can be written as
\bea
\sum {\Fmc^{\oQ}_L}_i  =x, \quad \sum {\Fmc_R}_i^d=y, \quad \sum {\Fmc_R}^u_i=z, \quad \sum {\Fmc^{\oL}_L}_i=u, \quad \sum  {\Fmc_R}^e_i=w.
\label{eq:not_sums_amly}
\eea
These expressions were introduced in the reference above, in the context of the Green-Schwartz mechanism \cite{Green:1984sg} but can have two different kinds of solutions in this non-supersymmetric context where anomalies must cancel with the fermionic fields of the effective theory:
\bea
\label{eq:sol_amly}
z=x,\quad  u=y,\quad  w=x \ \rightarrow \ 2 A^3 = 2 A^2 = 5 A^1/3= 3 x+ y=0 && \quad (a)\nn\\
z=-2x-y,\quad u=-3x,\quad w=4x+y. \quad && \quad (b)
\eea
Once the specific charges of quarks are fixed, by phenomenologically requirements, one must then solve $A_F=0$ with the charges of the charged leptons, or vice-versa. The charges of Table IV of Appendix A satisfy this constraint.  The solution \eq{eq:sol_amly}-(a) is the one that has been used in the context of supersymmetric Flavour symmetry (FS) but of course there $x$ and $y$ can be different from zero and not necessarily $3x=-y$. In our case, however this last expression it is the only solution for \eq{eq:sol_amly}-(a). To obtain the hierarchy of fermion masses using these constraints, let us define
 $p={\Fmc^{\oQ}_L}_1 + {{\Fmc_R}^u_2}$, $t={\Fmc^{\oQ}_L}_2 + {{\Fmc_R}^u_3}$ which are strongly constrained by the current value of fermion masses to be approximately $6$ and $2$ respectively, if the parameter expansion giving rise to the hierarchies of masses and mixing is of the order of the Cabibbo angle. Then the matrices of charges in the quark sector are as follows:
\bea
\label{eq:par_chr_anmly}
&&
\mathcal{C}^u=
\left[
\begin{array}{ccc}
\frac{1}{2} [p-t-3 {\Fmc_R}^u_2 -3 x - 2y] & p       &\frac{1}{2}[3 p +  t - 3 {\Fmc_R}^u_2 - x] \\
-p - x - y  \ & \frac{1}{2}[-p+t+3 {\Fmc_R}^u_2 +x]  &\ t \\
 -p -t - x - y \ & \frac{1}{2}[-p-t+3 {\Fmc_R}^u_2 +x]  &\ 0
\end{array}
\right],\nn\\
&&\mathcal{C}^d=
\left[
\begin{array}{ccc}
\mathcal{C}^d_{11} \ & {\Fmc_R}^d_2+p-{\Fmc_R}^u_2 & {\Fmc_R}^d_3+p-{\Fmc_R}^u_2\\
\mathcal{C}^d_{21} \ & \frac{1}{2}[2\,{{\Fmc_R}^d_2} - p + t +{{\Fmc_R}^u_2} + x] & \ \frac{1}{2}[2\,{{\Fmc_R}^d_3} - p + t + {{\Fmc_R}^u_2} + x] \\
\mathcal{C}^d_{31} \ & \frac{1}{2}[2\,{{\Fmc_R}^d_2} - p - t +
{{\Fmc_R}^u_2} + x] & \ \frac{1}{2}[2\,{{\Fmc_R}^d_3} - p - t +
{{\Fmc_R}^u_2} + x],
\end{array}
\right]\!\!,\nonumber\\
&&\mathcal{C}^d_{11} = -{\Fmc_R}^d_2-{\Fmc_R}^d_3 + p -{\Fmc_R}^u_2 -3x,\quad
\mathcal{C}^d_{21} = \frac{1}{2}[-2 \left({\Fmc_R}^d_2+{\Fmc_R}^d_3 \right)-p+t+{\Fmc_R}^u_2 +x +2 y]\nn\\
&& \mathcal{C}^d_{31} = \frac{1}{2}[-2 \left({\Fmc_R}^d_2+{\Fmc_R}^d_3 \right)-p+t-{\Fmc_R}^u_2 -x +2 y].
\eea
These parameterisations are valid for either solution of \eq{eq:sol_amly}. For \eq{eq:sol_amly}-b we have $3x\neq -y $, i.e. here they are independent parameters but for \eq{eq:sol_amly}-a $3x= -y$, and so $\mathcal{C}^{u,d}$ can be written entirely in terms of $x$ or $y$.

As a concrete example let us take this last class of solutions, i.e. \eq{eq:sol_amly}-a. Using as a constraint that the $(2,3)$ sectors of each matrix $\mathcal{C}^{u,d}$ must be positive and that $\mathcal{C}^d_{32}\geq$  $\mathcal{C}^d_{33}$ we have $-({\Fmc_R}^d_3+{\Fmc_R}^d_2)<{\Fmc_R}^u_3+x-p-t$ and using this we obtain $\mathcal{C}^d_{11}< -2 x-t$. On the other hand we have required $\mathcal{C}^u_{21}=-p + 2 x>0$ and, since $p$ and $t$ are positive, this implies that $0<p<2x$ and hence  $\mathcal{C}^d_{11}<0$.  In fact with this particular solution, i.e. requiring the sectors $(2,3)$ of both $\mathcal{C}^{u}$  and $\mathcal{C}^{d}$, the first column of the $d$ sector will contain only negative powers {\footnote{There are solutions with all elements positive for $\mathcal{C}^{u,d}$, however they do not correspond to a phenomenologically viable form of Yukawa matrices.}}.

 A way out of the problem just described, it is to allow for one $U(1)_F$, that we will call henceforward $U_{F_1}$, to generate the hierarchy of the sub-matrix mixing the second heaviest and the heaviest states in both quark sectors and use a discrete symmetry to forbid the operators 
$\overline{f_L}_i {f_R}_1$ $\! \Phi\left(\frac{v_{\varphi}/\sqrt{2}+\varphi}{\Lambda}\right)^{\qf\left({\Fmc_{f_L}}_i+{\Fmc_R^f}_1\right)}$. In this way all the negative contributions that could be associated to the sectors described by these operators are absent. Then another $U(1)$, $U(1)_{F_2}$,  will generate the structure giving rise to mixing between first and second generations and also to the masses of the lightest families.

In Table IV of the appendices 
we present an example of charges that can generate plausible Yukawa matrices. The structure generated by $U(1)_{F_1}\times Z_2$ it is as follows:
\bea
Y^u=\left[
\begin{array}{ccc}
\lambda^8 & \lambda^6 & \lambda^6\\
\lambda^6 & \lambda^4 & \lambda^2\\
\lambda^4 & \lambda^2  & 0
\end{array}
\right],\quad
Y^d=\left[
\begin{array}{ccc}
0 & \lambda^7 & \lambda^7\\
0 & \lambda^5 & \lambda^5\\
0 & \lambda^3  & \lambda^3
\end{array}
\right],\quad
Y^e=\left[
\begin{array}{ccc}
0 &  0 & 0 \\
0 & \lambda^4 & \lambda^3\\
0 & \lambda^3  & \lambda^2
\end{array}
\right],
\label{eq:yuk_varphi1}
\eea
where as mentioned before the first column of $Y^d$ is forbidden by the choices of the charges of the fields under $U(1)_{F_1}$ and the extra symmetry $Z_2$ and the elements of the first column and first row of $Y^e$ are forbidden by both, a combination on the choices of the charges of $U(1)_{F_1}\times Z_2$ and the appearence of fractional powers that cannot be present in renormalizable operators.
We illustrate the solutions of the type \eq{eq:sol_amly} -b for the second Abelian symmetry generated by  $U(1)_{F_2}$. The solution to the first three anomalies ($A_i$) of \eq{eq:anom_us_a} in terms of the parameters \eq{eq:sol_amly} is
\bea
\sum {\Fmc^{\oQ}_L}^\prime_i  =x', \quad \sum {\Fmc^d_R}_i^\prime=y', \quad \sum {\Fmc_R^u}^\prime_i=-2x'-y', \quad \sum {\Fmc^{\oL}_L}^\prime_i=-3x', \quad \sum  {\Fmc_R^e}^\prime_i=y'+4x',\nonumber \label{eq:pars_an_canc}\\
\eea
where the primed charges correspond to the SM fields as in Table 1, but for the second $U(1)_{F_2}$. The solution to $A_{F_2}=0$, completely analogous to the last expression of \eq{eq:anom_us_a}, requires in this case the inclusion of fermions beyond SM fermions, we will leave the discussion of these kind of solutions for a follow-up work. We basically want to generate a contribution just in the first column of d sector and since the u sector already exhibits a phenomenologically acceptable structure, we do not want to affect it too much. In order to be compatible with the $Z_2$ charges of Table IV 
we need to generate odd powers in the first column of $Y^d$. Taking into account all this, we can propose an easy solution where
\bea
{\Fmc^{\oQ}_L}^\prime_i  =0, \ \forall\ i\quad \Rightarrow x'=0 \quad
\Rightarrow \sum {\Fmc_R^u}^\prime_i=-y'.
\eea
If we further have ${\Fmc^{d}_R}^\prime_2={\Fmc^{d}_R}^\prime_3=0$ then we have that ${\Fmc^{d}_R}^\prime_1=y'$ must be odd. Due to the non zero sum of the charges ${\Fmc_R^u}^\prime_i$ we need to choose some $Z_2$ charges for the fields of the SM model which forbid the couplings
$\overline{Q}_1 \tilde \Phi u_j \left(\frac{\varphi_2}{\Lambda}\right)^{{\Fmc^{\oQ}_L}'_1+{\Fmc_R^u}'_j}$, where $\varphi_2$ is the flavon breaking the $U(1)_{F_2}$, and hence avoid a contribution to $Y^u$ from the flavon $\varphi_2$. In the lepton sector we want to generate a matrix element such that we have three non zero mass eigenvalues, keeping the conditions $\sum  {\Fmc_R^e}^\prime_i=y'$ odd and $\sum {\Fmc^{\oL}_L}^\prime_i=0$. There are many solutions to these equations but we choose the solution presented in Table IV 
because it is the one that gives the lightest gauge boson generating the $U(1)_{F_2}$. Then the contributions from $U(1)_{F_2}$ to the Yukawa matrix is zero for the up sector and has the following structure for down quark and charged lepton sectors:
\bea
Y^d=\left[
\begin{array}{ccc}
\lambda^{\prime 7} & 0 & 0 \\
\lambda^{\prime 7} & 0  & 0\\
\lambda^{\prime 7} & 0  & 0
\end{array}
\right],\quad
Y^e=\left[
\begin{array}{ccc}
\lambda^{\prime 7} &  0 & \lambda^{\prime 5} \\
0 & 0 & 0\\
0 & 0 & 0
\end{array}
\right].
\label{eq:yuk_varphi2}
\eea
Here we have called $\lambda'=\frac{\langle \varphi_2 \rangle}{M_2}$ and recall that $\lambda=\frac{\langle \varphi_1 \rangle}{M_1}$. In principle $M_1$ and $M_2$ could be close to each other, but as we we will see from the LEP bounds on $Z'$ bosons $M_1 > M_2$ but we can always have $\lambda'=\lambda$. The sum of the contributions of \eq{eq:yuk_varphi1} and  \eq{eq:yuk_varphi2} gives an appropriate description of masses for quarks and charged leptons and mixing in the quark sector. In the lepton sector the mixing must come from physics appearing beyond the scales $M_{1,2}$.

Of course the inclusion of a second $U(1)$ group in the set-up induces mixed anomalies between this and the first $U(1)$ and also among these and the hypercharge $U(1)_Y$ group. Let us enumerate them
\bea
\hat A_{F_2} &=&\text{Tr}\left[T^{U(1)_Y}_a\right] \text{Tr}\left[\{T^{U(1)_{F_1}}_b, T^{U(1)_{F_2}}_c  \}\right],\label{eq:an_AF2} \\
\hat A^\prime_{1} &=& \text{Tr}\left[T^{U(1)_{F_2}}_a \right] \text{Tr} \left[\{  T^{U(1)_{Y}}_b, T^{U(1)_{Y}}_c \} \right],\label{eq:ap_A1}\\
A_{{F_2} {F_1} {F_1}  }&=& \text{Tr}\left[ T^{U(1)_{F_2}}_a \{ T^{U(1)_{F_1}}_b,T^{U(1)_{F_1}}_c \}     \right],\label{eq:an_AF2F1F1}\\
\hat A_{{F_2} {F_1} {F_1}  }&=& \text{Tr}\left[ T^{U(1)_{F_2}}_a \right] \text{Tr} \left[ \{ T^{U(1)_{F_1}}_b,T^{U(1)_{F_1}}_c \}     \right],\label{eq:htan_AF2F1F1}\\
A_{{F_1} {F_2} {F_2}  }&=& \text{Tr}\left[ T^{U(1)_{F_1}}_a \{ T^{U(1)_{F_2}}_b,T^{U(1)_{F_2}}_c \}     \right],\label{eq:an_AF1F2F2} \\
\hat A_{{F_1} {F_2} {F_2}  }&=& \text{Tr}\left[ T^{U(1)_{F_1}}_a\right]\text{Tr}\left[ \{ T^{U(1)_{F_2}}_b,T^{U(1)_{F_2}}_c \}     \right].\label{eq:htan_AF1F2F2}
\eea
The anomaly $\hat A_{F_2}$, and the analogous $\hat A_{F_1}$ for the first family symmetry, are cancelled simply because with the fermions of the SM ~ $\sum_m T^{U(1)_Y}_a=0$. Also we need to satisfy $\text{Tr}\left[T^{U(1)_{F_2}}_a\right]$ and hence anomalies of Eqs. (\ref{eq:ap_A1}) and (\ref{eq:an_AF2F1F1}) cancel once this condition is satisfied. There is also an analogous anomaly to the one in \eq{eq:ap_A1} for the first $U(1)_{F_1}$:
\bea
\hat A_{1} &=& \text{Tr}\left[T^{U(1)_{F_1}}_a \right] \text{Tr} \left[\{  T^{U(1)_{Y}}_b, T^{U(1)_{Y}}_c \} \right],\label{eq:a_A1}
\eea
 but of course this and \eq{eq:htan_AF1F2F2} cancels once  $\text{Tr}\left[T^{U(1)_Y}_a\right]=0$. With the charges of Table IV of Appendix A, the equations $\text{Tr}\left[T^{U(1)_{F_{1,2}}}_a\right]=0$ are not satisfied with just the SM fermions. Hence we need to add other fermions charged only under  $U(1)_{F_{1,2}}$ and not under any of the SM gauge groups. Then Eqs. (\ref{eq:an_AF2F1F1}-\ref{eq:htan_AF1F2F2}) involve only these extra fields and do not affect the solution to the first three equations of (\ref{eq:anom_us_a}) and the analogous ones for the group factor $U(1)_{F_2}$.  We have checked that there are solutions of this type and we will present the results in a follow-up work.

\subsubsection{Gauge and scalar bosons.}

 The flavons $\varphi_1$ and $\varphi_2$ are singlets of the SM and have unitary $Z_2$ charge, then the most general scalar potential we can write down is
\bea
V &=&-\mu^2 \left |\Phi \right|^2+\lambda \left| \Phi \right|^4 -\mu_1^2 \left |\varphi_1 \right|^2+\lambda_{\varphi_1} \left| \varphi_1 \right|^4
-\mu_2^2 \left |\varphi_2 \right|^2+\lambda_{\varphi_2} \left| \varphi_2 \right|^4\nn\\
&& +2\left(\lambda'_1   \left |\Phi_1 \right|^2 \left |\varphi_1 \right|^2 + \lambda'_2   \left |\Phi_1 \right|^2 \left |\varphi_2 \right|^2  + \lambda'_{12}   \left |\varphi_1 \right|^2 \left |\varphi_2 \right|^2\right),
\eea
where $\Phi$ is the field representing the SM Higgs and the flavons can be parameterised as $\varphi_i=(v_{\varphi_i}+ h_i + ia_i)/\sqrt{2}$. It is easy to write down the minimization conditions:
\bea
\mu^2&=&\lambda v^2 + 2\left(\lambda'_1 v_{\varphi_1}^2+ \lambda'_2 v_{\varphi_2}^2\right)\nn \\
\mu^2_i&=&\lambda_i v^2_i + 2\left(\lambda'_i v^2+ \lambda'_{12} v_{\varphi_j}^2\right),\ j\neq i,\ i=1,2,
\eea
where $v/\sqrt{2},\  v_{\varphi_1}/\sqrt{2}\ $ and $v_{\varphi_2}/\sqrt{2}\ $ are respectively the vacuum expectation values of the Higgs and the flavons $\varphi_1$ and $\varphi_2$. The tree level squared mass matrix of these scalars can be obtained from $\partial^2 \langle V \rangle /\partial v_x \partial v_y $:
\bea
{\mathcal{M}}^2=2\left[
\begin{array}{ccc}
\lambda v^2     & \lambda'_2 v_{\varphi_2} v & \lambda'_1 v_{\varphi_1} v\\
\lambda'_2 v_{\varphi_2} v  & \lambda_{\varphi_2} v_{\varphi_2}^2  & \lambda'_{12} v_{\varphi_1} v_{\varphi_2}\\
\lambda'_1 v_{\varphi_1} v  & \lambda'_{12} v_{\varphi_1} v_{\varphi_2} & \lambda_{\varphi_1} v_{\varphi_1}^2
\end{array}
\right].
\eea
Considering the mixing between flavons and Higgs small and $v_{\varphi_1} \gg v_{\varphi_2}\gg v$ the squared mass eigenvalues are
\bea
&& m^2_H=2\lambda v^2\left(1-\frac{\lambda^{\prime 2}_2}{\lambda \lambda_{\varphi_2}} \right),\quad
m^2_{\varphi_2}= 2\lambda_{\varphi_2} v_{\varphi_2}^2\left(1+\frac{\lambda^{\prime 2}_2}{\lambda^2}\frac{v^2}{v_{\varphi_2}^2} \right),\nn\\
&& m^2_{\varphi_1}= 2 \lambda_{\varphi_1} v_{\varphi_1}^2\left(1+ \frac{\lambda^{'2}_{12}}{\lambda_{\varphi_1} \lambda_{\varphi_2}}\frac{v_{\varphi_2}^2}{v_{\varphi_1}^2}\right).
\eea
The Lagrangian of these scalars is then
\bea
{\mathcal{L}}= \left(D^\mu \Phi\right)^\dagger \left(D_\mu \Phi\right) + \left(D^\mu \varphi_1\right)^\dagger \left(D_\mu \varphi_1\right) + \left(D^\mu \varphi_2\right)^\dagger \left(D_\mu \varphi_2\right) - V,
\eea
where $D_\mu \Phi$ is just as in the SM and 
\bea
D_\mu \varphi_2&=&\frac{1}{\sqrt{2}}\left(\partial_\mu  -\frac{i}{2}g_{F_2} \qf_2 Z'_\mu  \right)\varphi_2 \nn \\
D_\mu \varphi_1&=&\frac{1}{\sqrt{2}}\left(\partial_\mu  -\frac{i}{2}g_{F_1} \qf_1 \tilde{Z}_\mu \right)\varphi_1,
\eea
since the hypercharge of the flavons is zero and the charge of $\varphi_2$ under ${U(1)_{F_1}}$ is zero and  the charge of $\varphi_1$ under ${U(1)_{F_2}}$ is also zero. Consequently, at tree level, there is no mixing among these gauge bosons and their masses are
\bea
m^2_Z   = \frac{1}{4}\frac{g^2}{\cos^2 \theta_w}v^2,\quad
m^2_{Z'}= \frac{1}{4} g^2_{F_2} \qf_2^2 v^2_2,\quad
m^2_{\tilde Z}= \frac{1}{4} g^2_{F_1} \qf_1^2 v^2_1.
\eea
The bounds from LEP \cite{:2003ih} can be used by identifying the contact interactions \cite{Eichten:1983hw} mediated by the extra Abelian gauge boson. These are basically the contribution to the amplitude of the process $\overline{e}e\rightarrow \overline{f} f$ process from the $s$ channel, mediated by a $Z'$ boson. Here the heaviest mass eigenstate is $Z_{1\mu}$, which is practically $\tilde Z$. On the other hand $Z_{2\mu}\sim Z'$ will be our lightest extra gauge boson. The effective relevant interaction for these processes is described by
\bea
\frac{\pm 4\pi}{(1+\delta_{ef}) \left(\Lambda^{f\pm}_{AB}\right)^2}\left(\overline{e}\gamma_\mu \Po_A e \right)\left(\overline{f}\gamma^\mu \Po_B f \right),
\label{eq:contact_zprime}
\eea
here $\Po_{A,B}$ label the left and right chirality projection operators, $\delta_{ef}=1,0$: it takes the value 1 for $e=f$ because there is an additional contribution from the $t$ channel and it is zero for everything else. The analogous Lagrangian to the neutral current interactions of the SM can be written as
$
{\mathcal{L}}= \sum_f \Fmc^f \overline{f} \gamma^\mu Z_{1\mu} f
$, hence the effective amplitude for the contact interactions, in the limit of the squared mass of the extra gauge boson is much higher than the $s$ parameter, it is:
\bea
\frac{g^2_F \overline{e} \gamma^\mu\left(\Fmc^e_{L}\Po_L +  \Fmc^e_{R}\Po_R  \right)e \overline{f} \gamma_\mu\left(\Fmc^f_{L}\Po_L + \Fmc^f_{R}\Po_R \right)f}{-m^2_{Z_1}}.
\eea
A bound on $m^2_{Z_1}/g^2_F$ can thus be easily  obtained:
\bea
m^2_{Z_1}\geq \frac{g^2_F}{4\pi}|\Fmc^e_{A} \Fmc^f_{B}|\left(1+\delta_{ef} \right)\left(\Lambda^{f\pm}_{AB}  \right)^2.
\label{eq:boundmz}
\eea
 Since the charges of different families are different for each kind of fermion, we have to use the specific decay channels (e.g. $e^+ e^- \to \mu^+ \mu^-$) presented in \cite{:2003ih} and not those for the combined leptonic and quark decays, which were obtained with the assumption that the charges of different families are the same. The most stringent bound comes from $e^+ e^- \to e^+ e^-$.  For example, for a pure vectorial contact interaction this would be the strongest limit with $\Lambda^{+}(V) > 21.7$ TeV. In order to make a rough estimate of how \eq{eq:contact_zprime} translates into a bound for our lightest extra gauge boson, $Z_{1\mu}$, we just have to input the charge of $e$ into \eq{eq:boundmz} for  $\Lambda^{\pm}_{A,B}$ with different combinations for $\{A,B\}=\{L,R\}$, since we have different charges for different chirality fermions, from Table $8.13$ of \cite{:2003ih} and then find the most stringent bound. We have then that $M_{1}/g_F\geq 26$ TeV. Using the same approach, we can have an idea of the scale of the ratio of the heaviest extra gauge boson to its coupling, coming from $U(1)_{F_1}$ by applying \eq{eq:boundmz}, since the charges of $e$ under this gauge boson are bigger we have then naturally a higher scale $M_{Z_2}/g_F\geq 116$ TeV. We can think that the couplings $g_{F_{1,2}}\ll 1$ and hence expect $M_{Z_1}$ and $M_{Z_2}$ at the TeV scale. However most part of the phenomenology depends on the inverse of this ratio, as it can be seen in $\S'$s \ref{sec:fnfc} and \ref{sec:prod} for the ``global'' case that we present there. Therefore, unlike models were fermions of the same type but of different $U(1)$ charges, it is not possible to lower down the scale of flavour changing interactions to few TeVs. However we can consider scenarios where only the heaviest fermions are coupled to an extra $Z'$ gauge boson and the others fermions to heavier ones, we will explore this in a subsequent work.
%
%
\section{Analysis with one flavon \label{subs:MIA_1_flavon}}

We consider some basic assumptions about the form of Yukawa matrices from experimental inputs and hierarchies that can be obtained with the FN mechanism. 
For simplicity  we focus on the $2\times 2$ sub matrix of Yukawa matrices which can be parameterised by
\begin{align}
Y=U_L^\dag {\widehat Y} U_R=
\sqrt2\frac{m_2}{v}\tc_L^{}\tc_R^{}
\left[\begin{array}{ccc}
\frac{m_1}{m_2}+\ttg_L^{}\ttg_R^{}e^{i(\phi_L-\phi_R)}&
-\frac{m_1}{m_2}\ttg_R^{}e^{i\phi_R}+\ttg_L^{}e^{i\phi_L}\\
-\frac{m_1}{m_2}\ttg_L^{}e^{-i\phi_L}+\ttg_R^{}e^{-i\phi_R}&
\frac{m_1}{m_2}\ttg_L^{}\ttg_R^{}e^{-i(\phi_L-\phi_R)}+1&
\end{array}\right],
\end{align}
where $m_i\,(i=1,2)$ are the mass eigenvalues of fermions, $U_X\,(X=L, R)$ are mixing matrices with ${U_X}_{12}=-\ts_X^{}e^{i\phi_X}=\sin\theta_Xe^{i\phi_X}$ and  ${U_X}_{11}=\tc_X$, so that $\ttg_X=\tc_X/\ts_X$.

As we have seen in $\S$ \ref{subsec:globalu1}, the FN mechanism can naturally generate hierarchical structures of Yukawa matrices, i.e., $Y_{11}<(Y_{12},Y_{21})<Y_{22}$ by assuming appropriate $U(1)$ charges. This means that the left- and right-mixing angles can be small and can be approximately written as $\ttg_{L/R}^{} \sim Y_{12/21}/Y_{22}\lesssim 1$.

The elements of the Yukawa matrices, $Y=U_L^\dagger Y_{\rm diag}U_R$ , can be expressed in terms of the rotation angles and hence we can identify its structure with that of a broken symmetry. What it is important in the analysis is to study the interplay between the contributions coming from the $u$ and $d$ sectors to the angles of the Cabibbo-Kobayashi-Maskawa matrix, $V_{\rm{CKM}}=U^u_L{U^d_L}^\dagger$.
For example, if the Cabibbo angle is given by $\lambda\sim\theta_L^u-\theta_L^d$ for small both left- and right-mixing angles, then without fine-tuning each $\theta^{u,d}_{12,L}$ would be at most of ${\mathcal{O}}(\lambda)$. Requiring $Y_{11}\sim \sqrt2m_1/v$ (stability of $Y_{11}$ under rotations), we then obtain the milder constraint on the right mixing angle as $m_1/m_2\gtrsim t_L^{}t_R^{}$. This can be generalized to three families and this has been widely explored in the literature \cite{Hall:1993ni}.

For three families, if some family symmetry forces $Y_{11}=0$, then the mass eigenvalues are related each other by  $m_1\sim m_2\ts_L^{12}\ts_R^{12}$, where now $m_1$ and $m_2$ correspond to the two lightest fermion masses. It is evident that if this relation is obtained for a $2$ by $2$ matrix, with $Y_{12}\sim Y_{21}$ then we have the so called  Gatto-Sartori-Tonin relation (GST) \cite{Gatto:1968ss} relation  $\ts_L\sim \ts_R\sim \sqrt{m_1/m_2}$. For symmetries involving three families for fermions which satisfy the GST relation, we have \footnote{{We also consider here as the ``GST relation'' the following expression with the correction factor $\sqrt{\frac{m_s}{m_d+m_s}}$ \cite{Kim:2004ki}.}} :
\bea
V_{us}=\left|\sqrt{\frac{m_d}{m_s}}-e^{i\Phi_a}\sqrt{\frac{m_u}{m_c}} \right|.
\label{eq:GST}
\eea
This relation is obtained by requiring: (a)~ the element $Y_{11}$ strongly suppressed, basically that $|Y_{11}|\ll |Y_{12}Y_{21}|/|Y_{22}|$, (b) $|Y_{12}|$ $\sim |Y_{21}|$, and (c) $s_{13}\ll s^u_{12}s^d_{23}, |s^d_{12}s^d_{23}|$. Then we can have $\text{s}^d_{12}=\sqrt{\frac{m_d}{m_s}}$, which it is $O(\lambda)$ and $\text{s}^u_{12}=\sqrt{\frac{m_u}{m_c}}$ which provides a small but relevant correction, in the light of the current precision measurements of $V_{us}$. Conditions (b) and (c) can be slightly altered and in this case we can have bigger contributions to $V_{us}$ from $\sqrt{\frac{m_u}{m_c}}$ in the form of an enhancement factor \cite{Masina:2006ad}.
The \eq{eq:GST} implies a relation between the quark masses and the mixing angles which is motivated by family symmetries.

For the lepton sector, the bi-large mixing angles of neutrinos can not be achieved by the hierarchical Yukawa coupling of charged leptons. It would be solved by further extensions of these models, for instance, by the introduction of TeV scale right-handed neutrinos. Such possibility is beyond the scope of this paper. Here we just simply assume hierarchical Yukawa matrices for quarks and charged leptons.

\subsection{Quark Sector \label{subs:oneflavonquarks}}

%

%
%

Given the order of magnitude of the $\ckm$ parameters
\bea
\left[\ckm\right]^*_{cd}= -\lambda, \quad
\left[\ckm\right]^*_{ts}= -A \lambda^2,\quad
\left[\ckm\right]^*_{td}=  A \lambda^3(1-\rho+i\eta),
\eea
we can express the mixing angles of both sectors as an expansion in terms of the parameter $\lambda$ as follows:
\bea
\ts^f_{12}= B^f_{12} \lambda + C^f_{12} \lambda^2 + D^f_{12} \lambda^3\nn\\
\ts^f_{13}= D^f_{13} \lambda^3 +    E^f_{13} \lambda^4 +    F^f_{13} \lambda^5  \nn\\
\ts^f_{23}= C^f_{23} \lambda^2 + D^f_{23} \lambda^3 + E^f_{23} \lambda^4.
\label{eq:param}
\eea
By unitary conditions on the diagonalising matrices, we can obtain relations between the $O(1)$ coefficients $B^u_{ij},\hdots, F^u_{ij}$ and those of $B^d_{ij},\hdots, F^d_{ij}$ and between these coefficients and the parameters $A$, $\rho$ and $\eta$ of the Wolfenstein parameterisation of the CKM matrix.  We can use the simplified form of the diagonalising matrices of each Yukawa matrix, \eq{eq:simdiagm}, in order to determine the relations among these coefficients and the phases appearing in \eq{eq:simdiagm}. The relations among  coefficients  in the $u$ and $d$ sectors are
\bea
&& B^u_{12}=1+B^d_{12}\nn\\
&& D^u_{23}=D^d_{23}\nn\\
&& C^u_{12}=C^d_{12}\nn\\
&& E^u_{13}\cos\phi_u=D^d_{13} \cos\phi_d-D^d_{23}=E^d_{13} \frac{\sin\phi_d}{\tan\phi_u}\nn\\
&& B^d_{12} B^u_{12} =2\left(D^u_{12}-D^d_{12}\right),
\eea
while the constraints coming from the $\ckm$ matrix are
\bea
&& C^u_{23}=A+C^d_{23}\nn\\
&& A\rho=-C^d_{23}+ D^u_{13} \cos\phi_u- D^d_{13} \cos\phi_d\nn\\
&& A\eta=-D^d_{13} \sin\phi_d + D^u_{13} \sin\phi_u.
\eea
We perform a numerical analysis for these cases by taking random inputs for the parameters in the $d$ sector and then determine those of the $u$ sector, whenever we cannot determine uniquely one parameter in the $u$ sector from the given $d$ parameters we take one of the solutions. When computing the branching ratios of the flavour violating process for this case we use the full numeric $\kappa^f_{ij}$ matrices, however it is easy to figure out the leading contributions:
\bea
\!\!\!&&\kappa^u=\nn\\
\!\!\!&&\!\!
\!\!\!\!\left[\!\!\!\! \begin{array}{ccc}
\begin{array}{c}
m_u p^u_{11}\\ +\lambda^2  m_c  (1+B^d_{12})^2 (p^u_{11}-2p^u_{12}+p^u_{22}) \\
+ \lambda^6 m_t O(1)
\end{array}
&
\begin{array}{c}
-\lambda m_c (1+B^d_{12}) (p^u_{12}-p^u_{22}) \\
+ \lambda^5 m_t O(1)
\end{array}
 &
-\lambda^3 m_t \text{e}^{-i\phi_u} D^u_{13} p^u_{13}\\
\hdots &
\begin{array}{c}
 m_c p^u_{22} \\
+ \lambda^4 m_t (A+C^d_{23})^2 (p^u_{22}-2p^u_{23})
\end{array}
&
 -\lambda^2 m_t p^u_{23} (A+C^d_{23})\\
\hdots & \hdots &
\begin{array}{c}
- \lambda^2 m_t 2 p^u_{23} \left(A+C^d_{23}\right)^2 \\
- \lambda^4 m_c p^u_{22} \left(A+C^d_{23}\right)^2
\end{array}
\end{array}
\!\!\! \!\right].\nn\\
&&
\label{eq:ku_eqcontcase}
\eea

In our numerical analysis, we concentrate on the following two cases;
\begin{itemize}
\item Q1: Case where most part of the mixing are controlled by $s^d_{ij}$.
\item Q2: Case where both $u$ and $d$ sectors are significant.
\end{itemize}
and the GST case (see $\S$ III C). 
These expressions apply to both examples Q1 and Q2 presented in the Table II of Appendix A, the difference is on the different charges chosen for both of them.
The flavour violating matrix for the $d$-type quarks, has the same structure except the element $\kappa^d_{33}$, the difference is the hierarchy of masses for the different type of quarks:
\bea
\kappa^d_{33}=m_b (p^d_{33} + O(\lambda^2) ).
\eea
Not surprisingly the elements $\kappa^f_{12}$, $\kappa^f_{13}$, $\kappa^f_{23}$ are, respectively, of the order $\lambda m_2\sim \lambda^5 m_3$, $\lambda^3 m_3$ and $\lambda^2 m_3$ due to the parameterisation in \eq{eq:param} which corresponds to the orders of the CKM mixing elements $\ts_{12}$, $\ts_{13}$ and $\ts_{23}$, respectively.

\subsection{Lepton sector \label{subsc:leptsect}}
We explore here the possibility that the mixing in the lepton sector comes from the charged leptons. In our notation the Pontecorvo-Maki-Nakagawa-Sakita matrix becomes
\bea
U_\text{PMNS}={U_L^e}{U_L^\nu}^\dag,
\eea
then $U^e_L$ is entirely determined by the mixing parameters measured in neutrino oscillations. In this case we assume then that the operator determining the neutrino masses is flavour diagonal. Given the current measurements and bounds of the mixing in $U_\text{PMNS}$, \cite{GonzalezGarcia:2008ru}, we can parameterise it in terms of the angles, \cite{Boudjemaa:2008jf}:
\bea
\ts_{13}= \frac{r}{{\sqrt{2}}},\quad
\ts_{12}=\frac{1 + s}{{\sqrt{3}}},\quad
\ts_{23}=\frac{1 + a}{{\sqrt{2}}}
\eea
and the phase $\delta$. The allowed values of these parameters are
\bea
0< r <0.22,\quad -0.12 <a <0.13,\quad -0.11< 0.04
\eea
and the form of $U_{\text{PMNS}}$ is
\bea
U_\text{PMNS}=\left[
\begin{array}{ccc}
  \frac{2 - s}{{\sqrt{6}}}   &\frac{1 + s}{{\sqrt{3}}} &
   \frac{r \,e^{i \,{\delta}}}{{\sqrt{2}}  }  \\
   \frac{\left( 1 + a \right) \,e^{i \,{\delta}}\,r\,\left( -2 + s \right)  +
      2\,\left( -1 + a \right) \,\left( 1 + s \right) }{2\,{\sqrt{6}}} &
   \frac{\left( -1 + a \right) \,\left( -2 + s \right)  -
      \left( 1 + a \right) \,e^{i \,{\delta}}\,r\,\left( 1 + s \right) }{2\,{\sqrt{3}}} &
   \frac{1 + a}{{\sqrt{2}}} \\
\frac{\left( 1 - a \right) \,e^{i \,{\delta}}\,r\,\left( -2 + s \right)  +
      2\,\left( 1 + a \right) \,\left( 1 + s \right) }{2\,{\sqrt{6}}} &
   \frac{\left( 1 + a \right) \,\left( -2 + s \right)  +
      \left( -1 + a \right) \,e^{i \,{\delta}}\,r\,\left( 1 + s \right) }{2\,{\sqrt{3}}} &
  \frac{1 - a}{{\sqrt{2}}}
\end{array}
\right]+O(r^2,s^2).
\eea
Then the flavour violating operators in the lepton sector become $\kappa_{\ell r}=(U_\text{PMNS})_{\ell i}\ $  $p^e_{ij}\ \left(U^\dagger_\text{PMNS}M^e_{\text{diag}} \ U^e_R \right)_{ij} $ $(U^{e\dagger}_R)_{jr}$. Since we do not have information from the right sector we can take two contrasting cases:
\begin{itemize}
 \item  L1: $U^e_R=U^*_\text{PMNS}$.
 \item  L2: $U^e_R=\mathbf{1}$.
\end{itemize}
It is difficult to study a general case of $U_R$ and we have just chosen a representative set of alternatives. The first one corresponds to a symmetric matrix form, motivated by its simplicity and predictivity. The second case is an extreme case where one expects mixings only from $U_\text{PMNS}$. According to \eq{Eq:k_ij} in general the flavour violating parameters $\kappa^e_{ij}$ receive a contribution from $U_R$ that is proportional to $m_i$.

\subsubsection{Case of ${U^e_R}=\mathbf{U^*_\text{PMNS}}$ ~ (L1) }
In this first case the Yukawa matrix of the charged lepton sector has a {\it democratic} structure:
\bea
&&Y^e=\nn\\
&&\left[
\begin{array}{ccc}
\frac{y_\tau}{6}\left(1- 6 r e^{-i\phi} \right)
&
\frac{y_\tau}{3\sqrt{2}}\left(-1+ \frac{r}{2}  e^{-i\phi} \right)&
\frac{y_\tau}{2\sqrt{3}}\left(1 - \sqrt{2} r  e^{-i\phi}  \right)\\
\hdots  & \frac{y_\tau }{3}\left(1+  r e^{-i\phi}\right) & \frac{y_\tau}{\sqrt{6}}\left(-1+\frac{r}{\sqrt{2}} e^{-i\phi} \right)\\
\hdots & \hdots & \frac{y_\tau}{2}
\end{array}
\right],
\eea
since all the elements of this matrix have a comparable contribution from $y_\tau$ and can be reproduced by a symmetry that gives a parameter expansion to the same power. Since what it matters in producing flavour violating parameters in $\kappa^e_{rs}$ are exactly the different powers in each of the entries of $Y^e$, in this case then all off-diagonal $\kappa^e_{rs}$ elements vanish. Of course we may consider that Yukawa matrices have the form
\bea
Y^e_{ij}=\lambda^{p_{ij}}+\lambda^{q_{ij}},
\eea
where the power $p_{ij}<q_{ij}$ controls the behavior of the Yukawa matrix leading to the appropriate mixing and eigenvalues, while $q_{ij}$ provides a substructure that can be used to explain the deviation from maximal mixing. However we are interested in associating the flavour violation with the leading contributions to the Yukawa matrices, hence we do not pursue further this possibility.
\subsubsection{Case of ${U^e_R=\mathbf{1}}$  ~ (L2)}

The case of $U^e_R=\mathbf{1}$ may be more interesting because depending on the value of the parameter $r$, it can produce different flavour violating elements. In this case the Yukawa matrix takes the form
\bea
Y^e=
\left[
\begin{array}{ccc}
\sqrt{\frac{2}{3}}y_e  \    &  \frac{y_{\mu}}{\sqrt{3}}    & \frac{re^{i\phi}}{\sqrt{2}} y_\tau \  \\
\frac{-y_{e}}{\sqrt{6}}(1+r e^{-i\phi})\  &  \frac{y_\mu}{\sqrt{3}}\left(1-\frac{re^{-i\phi}}{2} \right)\ & \frac{y_\tau}{\sqrt{2}}\  \\
\frac{y_e}{\sqrt{6}}(1-r e^{-i\phi}) & \frac{-y_\mu}{\sqrt{3}}\left(1+\frac{re^{-i\phi}}{2} \right)   & \frac{y_\tau}{\sqrt{2}}
\end{array}
\right].
\eea
Since $0<r<0.22$, the power $p^e_{31}$ is bigger than the rest of the powers in the first column of $Y^e$. All the elements of the second column are proportional to $y_\mu$ and the elements of the third one, proportional to $y_\tau$ with an $O(1)$ coefficient. Hence we can parameterise the powers of $Y^e$ as
\bea
\left[
\begin{array}{ccc}
p_1 & \ p_2 \ &\  p_3+n\\
p_1 & \ p_2 \ &\  p_3\\
p_1 & \ p_2 \ &\ p_3\\
\end{array}
\right],
\eea
where $n=1,2,..$ depending on the value of $r$. The flavour violating elements $\kappa_{\ell s}=(U_\text{PMNS})_{\ell i}\ p_{is} \ (U_\text{PMNS})^\dagger_{is}\ m_s$ take the form
\bea
\label{eq:krs_ur1}
\kappa_{\ell s}=
\begin{cases}
p_s m_s \delta_{\ell s}  & \text{for}\quad s=1,2,\ m_1=m_e,\ m_2=m_\mu\\
p_3 m_\tau \delta_{\ell 3} + n (U_\text{PMNS})_{\ell 1} (U_\text{PMNS})^*_{31} m_\tau  &\text{for}\quad s=3.
\end{cases}
\eea
We note that if $r$ takes its maximum value then the elements $Y^e_{21}$ and $Y^e_{31}$ could be suppressed by a power respect to the other element  $Y^e_{11}$ and then we can parameterise their powers as $p_1+1$ and $p_1+1$, respectively, and consequently their corresponding flavour violating elements as in \eq{eq:krs_ur1} with the appropriate replacements.
\subsection{GST scenario}
Here we choose to give as an input the size of the mixing angles in the up sector and hence to express the mixing angles of the down sector in terms of the measured CKM elements and the given up-quark mixing angles. When the later vanish we then basically identify the CKM mixing parameters to those of the down quark sector (again using the form of the standard CKM parameterisation for the diagonalising matrices and  $\ckm=U^u_L {U^d_L}^\dagger$):
\bea
\ts^{d}_{12} &=&\left|\left(\tc^u_{12} \tc^u_{23}-\text{e}^{i\phi_u} \ts^u_{12}  \ts^u_{13}  \ts^u_{23}\right)V^*_{cd} + \tc^{u}_{13} \ts^u_{12} V^*_{ud}-\left(\tc^u_{12} \ts^u_{23}-\tc^u_{23}\ts^u_{12}\ts^u_{13} \text{e}^{i\phi_u} \right) V^*_{td} \right|\nn\\
\ts^{d}_{13}\text{e}^{-i\phi_d} &=&  \tc^u_{13} \tc^u_{23}  V^*_{td}-\text{e}^{i\phi_u} \ts^u_{13} V^*_{ud} +\tc^u_{13} \ts^u_{23} V^*_{cd}\nn\\
\ts^{d}_{23} &=&\left|\tc^u_{13} \tc^u_{23} V^*_{ts} + \tc^u_{13} \ts^u_{23} V^*_{cs}+ \text{e}^{-i\phi_u}\ts^u_{13} V^*_{us}\right|
\eea
The FV matrices, $\kappa^\fsm$, of \eq{Eq:L_eff} can be written for these cases
\bea
\label{eq:fv_ks_msym}
\kappa^\fsm_{lk}=\frac{v}{\sqrt{2}} {U^{\fsm}_{L}}_{l i}\ \left[ p^\fsm_{ij} Y^\fsm_{ij}\right] \ {U^\fsm_{L}}^\T_{j k}
=
\kappa^\fsm_{lk}=\frac{v}{\sqrt{2}} {U^{\fsm}_{L}}_{l i}\ \left[ p^\fsm_{ij} \left( {U^\fsm}^\dagger Y^\fsm_{\rm{diag}} {U^\fsm}^* \right)_{ij}  \right] \ {U^\fsm_{L}}^\T_{j k}.
\eea
where we have taken $U_R=U^*_L$ and we have not split the power $p^\fsm_{ij}$ in terms of the sum of charges ${f^u_L}_i+{f^u_R}_j$, since at the end what it is constrained from reproducing the mixing angles and masses of quarks is the sum and not the individual charges. We can express the Yukawa matrices in terms of the assumed inputs for the up quarks and the CKM elements and hence we determine the size of the elements $\kappa^\fsm_{lk}$, the specific form of this result can be obtained from the expressions in the Appendix (\ref{ap:flvpars}).
As we have mentioned GST-like symmetries fix the order of magnitude of the diagonalising angles, such that
\bea
&&\ts^u_{12}\sim \sqrt{\frac{m_u}{m_c}},\quad \ts^u_{23}\sim \frac{m_c}{m_t},\quad \ts^u_{13}\sim \frac{m_u}{m_t},\nn\\
&&\ts^d_{12}\sim \sqrt{\frac{m_d}{m_s}},\quad \ts^d_{23}\sim \frac{m_s}{m_b},\quad \ts^d_{13}\sim \frac{m_d}{m_b},
\label{eq:gst_quarkrel}
\eea
hence the flavour violating matrices for both up and quark flavour violating parameters have a very particular form. Assuming the simplified version of the diagonalising matrices:
\bea
\label{eq:simdiagm}
U^T=\left[
\begin{array}{lll}
 1-\frac{\ts^2_{12}}{2} & \ts_{12} & e^{-i \text{$\phi $}} \ts_{13} \\
 -\ts_{12} & 1-\frac{\ts^2_{12}}{2} & \ts_{23} \\
 \ts_{12} \ts_{23}-e^{i \text{$\phi $}} \ts_{13} & -e^{i \text{$\phi $}} \ts_{12}
   \ts_{13}-\ts_{23} & 1
\end{array}
\right],
\eea
where in this case the elements of $U$ for each sector are given by
\bea
\begin{array}{l}
U^u_{12}\sim   \lambda^2\\
U^u_{23}\sim   \lambda^4\\
U^u_{13}\sim   e^{-i\phi\prime}\lambda^7
\end{array},\quad
\begin{array}{l}
U^d_{12}\simeq \left[\ckm\right]^*_{cd}= -\lambda \\
U^d_{23}\simeq \left[\ckm\right]^*_{ts}= -A \lambda^2\\
U^d_{13}\simeq \left[\ckm\right]^*_{td}=  A \lambda^3(1-\rho+i\eta).
\end{array}
\eea
Taking into account the current values of the quark masses, we have $Y^d_{\rm{diag}}\approx \text{Diagonal}\left\{\lambda^7,\lambda^5,\lambda^2 \right\}$ and $Y^u_{\rm{diag}}\approx \text{Diagonal}\left\{\lambda^7,\lambda^4,1 \right\}$ hence, just to have an idea in terms of the parameter $\lambda$, we have
\bea
&& \kappa^u\sim
\frac{v}{\sqrt{2}}
\left[
\begin{array}{ccc}
\lambda^7 &  \lambda^6 & \lambda^6 \\
... & \lambda^4 & \lambda^4 \\
... & ... &   \lambda^{12}
\end{array}
\right],\quad\quad
\kappa^d \sim
\frac{v}{\sqrt{2}}
\left[
\begin{array}{ccc}
 \lambda^7 &  \lambda^6  & \lesssim A \lambda^5\\
...          &  \lambda^5  &  \lambda^4 \\
...          &  ...          &  \lambda^2
\end{array}
\right]
\eea
Also in the GST-like models it is possible to determine the lepton flavour violation (LFV) matrices since here also the mixing angles are of the form
\bea
\ts^e_{12}=\sqrt{\frac{m_e}{m_\mu}},\quad \ts^e_{23}=a^e_{23} \frac{m_e}{m_\mu}\sim \frac{m_\mu}{m_\tau},\quad
\ts^e_{13}=a^e_{13} \frac{m_e}{m_\tau}\ll \left(\frac{m_e}{m_\tau}\right)^{1/2},
\label{eq:gst_leptsrel}
\eea
where the diagonalising matrix has the form of \eq{eq:simdiagm}. In these models the large mixing in the lepton sector must come from neutrino sector, which we do not explore here, and the charged LFV  matrix has the form
\bea
&&\kappa^e\sim
\frac{v}{\sqrt{2}}
\left[
\begin{array}{ccc}
\lambda^9 & \lambda^8 & \lambda^9\\
\hdots    & \lambda^6 & \lambda^6\\
\hdots    & \hdots    & \lambda^7
\end{array}
\right].
\label{eq:gst_ke}
\eea
For this case, there is not an obvious set of $U(1)$ charges that could reproduce entirely the structure of Eqs.~ (\ref{eq:gst_leptsrel}) and (\ref{eq:gst_quarkrel}), however in the effective analysis we can assume that only one flavon generates this structure in each sector and therefore we probe the implications of it from flavour changing processes.
%
%
%
%
%

%
\section{Flavour changing decays via the flavon \label{sec:fnfc} }
%
%
In this section we analyse the flavour changing decays via one flavon, based on the examples presented in $\S$ \ref{subs:MIA_1_flavon}. For the numerical analysis we first make a numerical fit of the Yukawa matrices that is in agreement with a choice of $U(1)$ charges (except for the GST case) and then extract the exact numerical form of $\kappa^f_{ij}$, obtained from the diagonalising matrices of the Yukawa couplings and the quark mass values, as given in \eq{Eq:k_ij}, and for simplicity we take all the elements of $\kappa^f$ to be real.

\begin{figure}[tb]
\centering
\includegraphics[width=6cm]{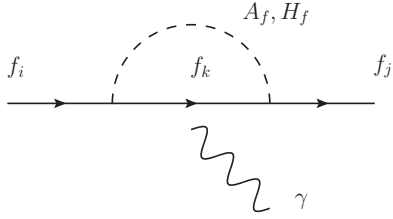}
\caption{The Feynman diagram responsible for the flavour changing decays of fermions through a flavon.} \label{FIG:fifjgamma}
\end{figure}

With the flavour changing couplings of \eq{Eq:L_eff} a photonic di-pole operator is induced at one-loop level. The corresponding effective interaction Lagrangian is described by
\begin{align}
{\mathcal L}_\text{di-pole} &= \overline{f}_i
({A_L^\gamma}_{ij}\text{P}_L+{A_R^\gamma}_{ij}\text{P}_R)
\sigma^{\mu\nu}f_jF_{\mu\nu}+\text{H.c.}
\quad(i\ne j), \label{Eq:L_fifjgamma}
\end{align}
where $\text{P}_{L/R}$ are the chiral projection operators and
$\sigma^{\mu\nu}=\frac{i}2[\gamma^\mu,\gamma^\nu]$. The
coefficients ${A_{L/R}^\gamma}_{ij}$ are calculated from the penguin diagram in Fig.~\ref{FIG:fifjgamma} and are given by
\begin{align}
{A_L^\gamma}_{ij}=\frac1{(4\pi)^2}\frac{Q_f\,\text{e}} {2v_\varphi^2}
 \left[- \kappa_{jk}^{f}\kappa_{ki}^f m_k\ c_{11} + \kappa_{jk}^{f}\kappa_{ik}^{*f} [m_j(c_{11}-c_{12}+c_{21}-c_{23})+m_i(c_{12}+c_{23})]\right],
\nonumber\\
{A_R^\gamma}_{ij}=\frac1{(4\pi)^2}\frac{Q_f\,\text{e}} {2v_\varphi^2}
 \left[- \kappa_{kj}^{*f}\kappa_{ik}^{*f} m_k \ c_{11} + \kappa_{kj}^{*f}\kappa_{ki}^{f} [m_j(c_{11}-c_{12}+c_{21}-c_{23})+m_i(c_{12}+c_{23})]\right],
\label{Eq:ALAR_gamma}
\end{align}
for a CP odd flavon, where the index $k$ denotes the internal fermions and $c_{ab}$ are the Passarino-Veltman functions \cite{Passarino:1978jh}, whose arguments are defined as $c_{ab}=C_{ab}(m_j^2,0,m_i^2,m_\varphi^2,m_k^2,m_k^2)$.  We note that  for a CP even flavon, $m_k$ is replaced by $-m_k$ in Eqs.~ (\ref{Eq:ALAR_gamma}),  we call such contributions ${A_{L,R}^{\gamma +}}_{ij}$. For values of $\lambda_\varphi v_\varphi^2$ comparable to $M^2$ both contributions are important. For gauged models, where we only have a CP even flavon which is quite heavy, at least $ v_\varphi > 26$ TeV in the cases studied here, the coefficients ${A_{R,L}^{\gamma +} }_{ij}$ would be highly suppressed in any case.

\begin{figure}[tb]
\centering
\includegraphics[width=10cm]{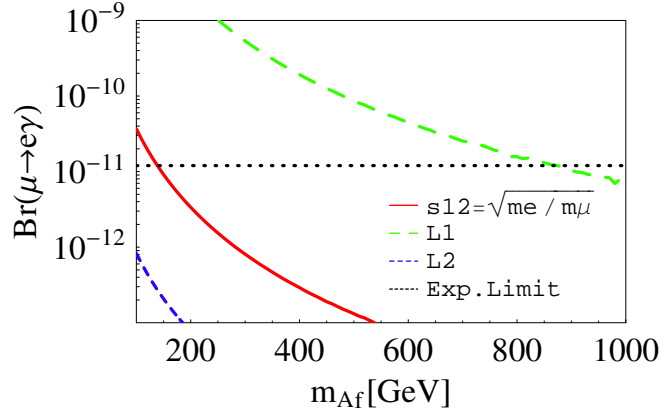}
\caption{The decay branching ratio of the process $\mu\to e\gamma$ via the CP odd flavon $A_f$, shown as a function of the flavon mass.
Solid, long-dashed and dashed curves denote the cases of $s_{12}^e=\sqrt{m_e/m_\mu}$, L1 and L2 with $v_\varphi=500$ GeV. The current experimental upper bound is also shown (horizontal dotted line).} \label{FIG:FN-MuEGamma}
\end{figure}

The most stringent bound on the flavon scenarios presented in $\S$ \ref{subs:MIA_1_flavon} may come from the lepton flavour violations $\ell_i\to\ell_j \gamma(i\ne j)$, whose partial decay widths are given by
\begin{align}
\Gamma_{\ell_i\to\ell_j\gamma}^{} &=
\frac{m_i^3}{4\pi}\left(1-\frac{m_j^2}{m_i^2}\right)^3
\left(\left|{A_L^\gamma}_{ij}\right|^2+\left|{A_R^\gamma}_{ij}\right|^2\right).
\label{Eq:G_fifjgamma}
\end{align}
In Fig.~\ref{FIG:FN-MuEGamma} we show the decay branching ratio of $\mu \to e\gamma$ for the examples where $\ts_{12}=\sqrt{m_e/m_\mu}$ (GST-like),  L1 for which where $U^e_L=U_{PMNS}$ and $U^e_R=U^*_{PMNS}$ and finally L2, where $U^e_L=U_{PMNS}$ and $U^e_R={\bf 1}$. For all of them we have assumed $v_\varphi=500$ GeV.
 For case L1, flavon masses up to 1 TeV are really disfavoured, as they produce a decay already above the current experimental bound, $B(\mu\to e\gamma)\leq 1.2 \times 10^{-11}$ at the $90$ \% C.L. \cite{exp:muegamma}, and could easily saturate its expected improvement \cite{Cattaneo:2009zz}. On the other hand, for the case L2, where $U^e_L=U_{PMNS}$ and $U^e_R=1$, relatively light flavons ($m_{A_f}\lesssim 150$ GeV) can be allowed under experimental constraints. However the bound obtained could be shifted by the detailed flavour structure and of course by the value of $v_\varphi$. We note that in the GST-like case, where $\ts_{12}=\sqrt{m_e/m_\mu}$ all the leptonic flavour changing violating processes are also suppressed, in particular the decay $\mu \to e\gamma$  would have a $\lambda^4$ suppression with respect to the L1, according to the flavour violating matrix $\kappa^e$ of \eq{eq:gst_ke}. Hence this case is still safely below the current bounds for masses $m_\varphi>150$ GeV.
 Since predictions of all these cases are quite sensitive to diagonalising matrices, the study of this decay is a good probe of this kind of scenarios. It is worth to note that the bound from $\mu\to e\gamma$ is relative stronger than other LFV decay modes such as $\mu\to 3e$ and $\tau\to\mu\gamma$. This is because the radiative LFV decay $\mu\to e\gamma$ can be enhanced by tau mass and tau LFV coupling, and the tree level decay is strongly suppressed by the electron mass.

We consider now the analogous decay processes for the top quark and also the processes $q_jg (q_j=u,c)$ and $q_j Z$.
Experimental searches for these FCNC decay modes have been analysed by the ZEUS and the CDF collaborations \cite{Chekanov:2003yt,:2008aaa}. The partial decay widths of these are given by
\begin{align}
\Gamma_{q_i\to q_jg}^{} &= \frac43
\frac{m_i^3}{4\pi}\left(1-\frac{m_j^2}{m_i^2}\right)^3
\left(\left|A_L^\gamma\right|^2+\left|A_R^\gamma\right|^2
\right)_{Q_f\text{e}\to g_S^{}},\\
\Gamma_{t\to q_jZ}^{} &=
\frac{m_t^3}{4\pi} \left(1-\frac{m_Z^2}{m_t^2}\right)^2
\left(1+\frac{2m_Z^2}{m_t^2}\right)
\left(\left|A_L^Z\right|^2+\left|A_R^Z\right|^2 \right),
\end{align}
with $m_j=0$, where
\begin{align}
&A_{L}^Z=\frac1{(4\pi)^2}\frac{\,\text{e}}{2v_\varphi^2} \left\{-\kappa_{jk}^{f}
\kappa_{k3}^fm_k\left[c_V^fc_{11}^Z-
c_A^f(c_{11}^Z-2c_{12}^Z)\right]+
\kappa_{jk}^{f}\kappa_{3k}^{f*}m_t(c_V^f-
c_A^f)(c_{12}^Z+c_{23}^Z)\right\},\\
&A_{R}^Z=\frac1{(4\pi)^2}\frac{\,\text{e}}{2v_\varphi^2} \left\{-\kappa_{jk}^{f*}
\kappa_{k3}^{f*}m_k\left[c_V^fc_{11}^Z-
c_A^f(c_{11}^Z-2c_{12}^Z)\right]+
\kappa_{kj}^{f*}\kappa_{k3}^fm_t(c_V^f-
c_A^f)(c_{12}^Z+c_{23}^Z)\right\},
\end{align}
for a CP odd flavon, and $c_V^f = \frac1{2s_Wc_W}\left(T_{3L}^f-2Q_fs_W^2\right), c_A^f
= \frac1{2s_Wc_W}\left(-T_{3L}^f\right)$. The Passarino-Veltman functions are replaced by their corresponding mass arguments as $c_{ab}^Z=C_{ab}(m_j^2=0,m_Z^2,m_i^2,m_\varphi^2,m_k^2,m_k^2)$\cite{Passarino:1978jh}.

A large top-quark mass in the SM is natural, therefore in our approach we decided to give examples of $U(1)_F$ charges for which the sum $\Fmc^{\oQ}_3+{\Fmc_R^u}_3=0$, which as a consequence produces a matrix element $\kappa_{tt}=\kappa^u_{33}$ very small (see e.g \eq{eq:ku_eqcontcase}). Due to this, the branching ratios of flavour changing decays of top quarks, such as $t\to cg,~ c \gamma, c Z$ are at most $\mathcal{O}(10^{-10})$ for case Q1 and Q2 (for the GST these are even more suppressed according to \eq{eq:gst_quarkrel}) and so these rare decay modes are out of the experimental reach both at the LHC and the ILC\cite{AguilarSaavedra:2000aj}.

\begin{figure}[tb]
\centering
\includegraphics[width=10cm]{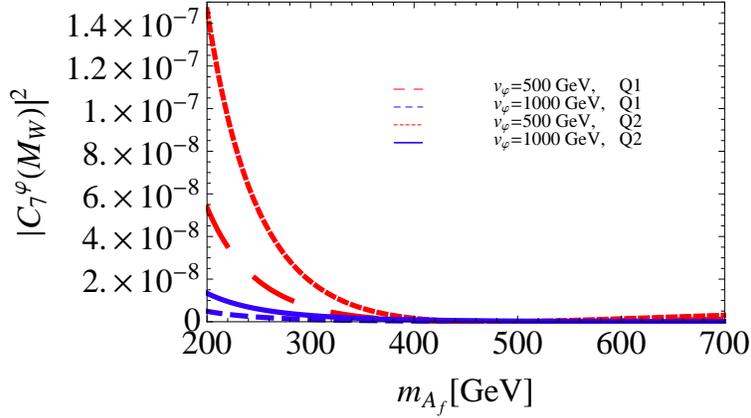}
\caption{The operator $|C^{\varphi}_7|^2$ at $M_W$. In the SM  $|C^{(0)eff}_7|^2$ at $M_W$ has a value of $0.036$.
Thus the contribution at this scale from flavon processes is quite tiny even for relative light values of $v_\varphi$ and $m_\phi$. For simplicity we have taken $\lambda_\varphi=1/2$. See text to check how this coefficient enters into the effective decay amplitude.} \label{FIG:C7gst}
\end{figure}

For the decay  $b\to s\gamma$ the flavour violating parameters $A_{L,R}^\gamma$  have the same form as in \eq{Eq:G_fifjgamma}. However since we have to add the contributions from the flavon to the SM contributions mediating  $b\to s\gamma$, it is customary to express it in terms of the Wilson Coefficients introduced in the effective Hamiltonian approach \cite{Gambino:2001ew,bsg:gen}:
$H^W_{eff}=\frac{-4 G_F}{\sqrt{2}}V_{tb}V^*_{ts} \sum^{8}_{i=1} C^W_i(\mu) O_i(\mu)$. 
 The most important contributions are those of $C_7$ and $C_8$, since the corresponding operators are defined as: ~
$O_7=\frac{e}{16\pi^2} m_b \bar{s} \sigma^{\mu\nu}b_R F_{\mu\nu},\quad O_8=\frac{e}{16\pi^2} m_b \bar{s} \sigma^{\mu\nu}T^a b_R G_{\mu\nu}$, which describe the emission of a photon and a gluon respectively. At the EW scale obviously only $O_7$ gives a contribution to $b\to s\gamma$, however at the decay scale $\mu_b$, the contributions from both $O_7$ and $O_8$ enter into the contribution of the decay due QCD corrections. Analogously to operators $O_7$ and $O_8$ we have the operators $O'_7$ and $O'_8$ with opposite chirality (i.e. $f_L \leftrightarrow f_R$) but with  Wilson coefficients, $C^\prime_i$, suppressed by the ratio $m_s/m_b$. In the SM $C_{7,8}$ are mediated by $W^-$. In our case we can define the effective Lagrangian $H^\phi$, such that $H_{eff}= H^W_{eff}+H^\phi_{eff}$, hence $H^\phi_{eff}(\mu)= \frac{-4 G_F}{\sqrt{2}}V_{tb}V^*_{ts} \sum^{8}_{i=1} \left[\frac{e}{16\pi^2}m_b\right] \bar{s} \left[C^\varphi_i(\mu) P_R+ C^{\varphi\prime}_i(\mu) P_L \right] b F_{\mu\nu}$, and comparing this to the effective Lagrangian of \eq{Eq:L_fifjgamma} we have then that
\bea
C^\varphi_7(\mu)&=&\frac{16\pi^2}{e}\frac{1}{4 G_F V_{tb}V^*_{ts}m_b}\left[ \hat A^\gamma_{L 32}(\mu)+ \hat A^{\gamma +}_{L 32}(\mu)  \right],\nn \\
C^{\varphi\prime}_7(\mu)&=&\frac{16\pi^2}{e}\frac{1}{4 G_F V_{tb}V^*_{ts}m_b} \left[ \hat A^\gamma_{R 32}(\mu) + \hat A^{\gamma +}_{R 32}(\mu) \right].
\eea
Calling $a^1_{L ij}$ to the first summand of $A_{L ij}^\gamma$ and $a^1_{R ij}$ to the analogous of $A_{R ij}^\gamma$ in \eq{Eq:ALAR_gamma}, we have that $\hat A_{(L,R) ij}^{\gamma}$ $=A_{(L,R) ij}^\gamma + a^{1*}_{(R,L) ij} m_s/m_b$, but from here the dominant term is always the first. We have the analogous relations for the coefficients  $\hat A_{(L,R) ij}^{\gamma +}$.  Since the effective $U(1)_F$ symmetry breaking scale is below 1 TeV, we can assume the form of $A_{L,R}^\gamma$ to be the same as that of \eq{Eq:ALAR_gamma}, which it is formally at the electroweak scale: i.e. $A_{L,R}^\gamma(M_{\text{F}})=A_{L,R}^\gamma(M_{W})$. Then one just would have to take care of the QCD corrections from $M_{W}$ to the decay scale $\mu_b$.  We do not calculate them here but we expect them to be small, unlike the SM ones that receive important contributions due to the top quark in the loop involved in the decay \cite{Gambino:2001ew}.
In Fig.~\ref{FIG:C7gst} we present a plot of $|C^\varphi_7(M_{W})|^2$ versus $m_{\varphi}$ for the cases Q1 and Q2 introduced in $\S$ \ref{subs:MIA_1_flavon}. We included here the contribution from the CP-even part because for the low values considered in Fig.~\ref{FIG:C7gst} both contributions are important and its only for small values of about $v_\varphi=150$ GeV (and $m_\varphi$ above 200 GeV) that we start having an enhancement to $|C^{\varphi}_7(M_W)|^2$,  $\mathcal{O}(10^{-2} \%)$, which nevertheless, it is really negligible. Smaller values than $v_\varphi=100$ GeV could have an impact at the 1 \% level but these values are not realistic within our framework. Hence this process practically does not give constraints for bounds on the values of $v_\varphi$ and $m_\varphi$.
 Since the QCD corrections of $C^\varphi_i(\mu)$ and $C^W_i(\mu)$ are different, the decay width of $b\to s \gamma$ at the leading correction in $\alpha_s$, can be expressed as
\begin{align}
\Gamma_{B\rightarrow X_s\gamma} = \frac{\alpha}{16\pi^4} G^2_F m_b^5 |V_{tb}V^*_{ts}|^2\left[\left|C^{\text{SM}} (\mu_b)\right|^2+\left|C^{\text{SM}\prime}(\mu_b)\right|^2   + \left|C^{\varphi} (\mu_b)\right|^2+\left|C^{\varphi\prime}(\mu_b)\right|^2    \right].\label{Eq:Gamma_bsg}
\end{align}
In the SM $|C^{(0)eff}_7(M_W)|^2=0.036$ \cite{Gambino:2001ew} and the QCD corrections bring this value up to $0.094$ at the decay scale $\mu_b$. Therefore the values considered cannot alter significantly the SM value of $B[b \rightarrow s \gamma]=(3.15\pm 0.23)\times 10^{-4}$ \cite{sm:bsg}. From \eq{Eq:Gamma_bsg} we see that in this scenario we have the extra contribution of $C^{\varphi\prime}(\mu_b)$ which however is of the same order of magnitude of $C^{\varphi}(\mu_b)$ in the Q2 case and suppressed for the Q1 example, hence it cannot alter either significantly the value of the SM $B[b \rightarrow s \gamma]$. The difference to its experimental value of $B[b \rightarrow s \gamma]=(3.55\pm 0.24^{+0.09}_{-0.10}\pm 0.03)\times 10^{-4}$ \cite{exp:bsg}, is the source of current debates on flavour models.

%
\section{Production and decays of the flavon at the LHC \label{sec:prod}}

In this section we discuss the flavon production mechanism and its decay patterns at the LHC for the examples presented in $\S$ \ref{subs:MIA_1_flavon}. For the numerical analysis we proceed as in $\S$ \ref{sec:fnfc}.

\begin{figure}[tb]
\centering
\includegraphics[width=9cm]{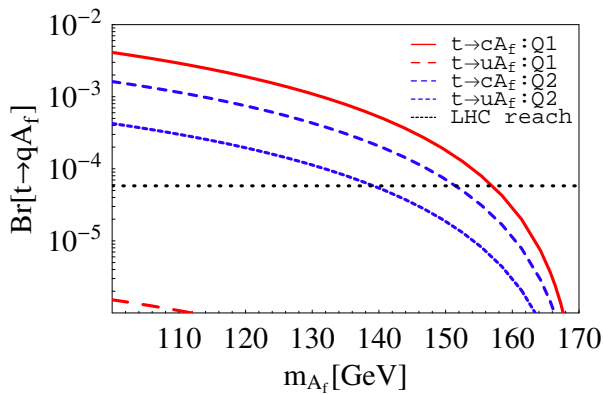}
\caption{The branching ratio of top quark flavour changing decays into flavons for examples of  $\S$ \ref{subs:MIA_1_flavon}. Solid and long-dashed curves denote, respectively, the decay $t\to cA_f$ and $t\to uA_f$ for case Q1. Those for the case Q2 are shown as dashed and short-dashed curves. For all these cases we have taken $v_\varphi=500$ GeV. The expected LHC reach for those FCNC decays is also shown (horizontal dotted line). }
\label{FIG:tcPhi}
\end{figure}

 In the models with anomalous FCNC top-quark coupling, the top-quark can decay into Higgs boson as $t\to ch$ \cite{Babu:1999me,Giudice:2008uua}. Similarly to the Higgs boson, flavons can also be produced in top-quark FCNC decay.
The decay widths at the tree level of the processes $t\to q_jA_f$, are given by
\begin{align}
\Gamma(t\to q_jA_f) = \frac{G_Fm_t}{4\sqrt2\pi}
\left(\frac{v}{v_\varphi}\right)^2 \left(\left|\kappa_{j3}^u\right|^2+\left|\kappa_{3j}^u\right|^2\right)
\left(1-\frac{m_\varphi^2}{m_t^2}\right)^2,
\end{align}
where $q_j$ is any of the other, than the top-quark, $u$ type SM fermions and $\kappa^u_{j3}$ the flavour violating parameters as defined in \eq{Eq:k_ij}. In Fig.~\ref{FIG:tcPhi} we show the decay branching ratios for the processes $t\to cA_f$ and $t\to uA_f$ for the cases Q1 and Q2 introduced in $\S$ \ref{subs:oneflavonquarks}.
The expected upper limits of 
such branching ratios 
at the LHC could reach up to ${\mathcal O}(10^{-5})$. The ILC would
improve that experimental precision up to ${\mathcal
O}(10^{-6})$\cite{AguilarSaavedra:2000aj}.
When $m_{A_f} \sim 150$ GeV and $v_\varphi=500$ GeV, we can see that for the case Q2, where the mixing angles in the $u$-sector are CKM-like \eq{eq:param}, both of the decays $t\to c{A_f}, u{A_f}$ could be tested at the LHC. This happens because $\kappa_{tc}=\kappa^u_{32} \sim \lambda^2 m_t$ while $\kappa_{tu}\sim \lambda^3 m_t$. The decay $t\to c{A_f}$ in case Q1 where mixing is mostly led by the $d$ sector $\kappa_{tc}$ is still at the order of $\lambda^2 m_t$, however $\kappa_{tu} \sim \lambda^4 m_t$, for this reason the decay $t\to u{A_f}$ is highly suppressed, and therefore it would only have a chance to be probed at the ILC. The GST case would also be suppressed as like case Q1.

\begin{figure}[tb]
\includegraphics[width=10cm]{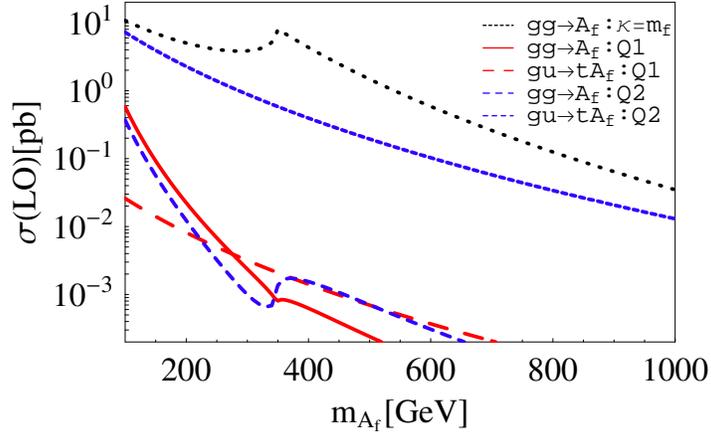}
\caption{
 The production cross sections of the gluon fusion and the FCNC single top production are shown as a function of the flavon mass. The center of mass energy for the $pp$ system is taken at $14$ TeV and we have chosen $v_\varphi=500$ GeV.
For case Q1 (Q2), $\sigma(gg\to A_f)$ is shown as a solid (dashed) curve, while $\sigma(gu\to tA_f)$ is shown as a long-dashed (short-dashed) curve. The case for $\kappa_{ff}=m_f$ is also shown for comparison.}
\label{FIG:GluonFusion}
\end{figure}

While for the SM Higgs, the vector boson fusion $VV^*\to H (V=W^-,Z)$ and the Higgs strahlung $q{\bar q}'\to VH$ are relevant for the Higgs production, in our case the flavon does not interact with gauge bosons at the tree level, therefore the analogous processes can not be used for flavon production. As it happens with the SM Higgs boson, we can expect that the main production channel of the flavon, when $\kappa_{tt}$ is sufficiently large, it would be the gluon fusion mechanism $gg\to A_f$ at a high energy Hadron collider\cite{Babu:1999me,Ref:ggh}. The production cross section of the flavon via this mechanism at the LHC is estimated in Fig.~\ref{FIG:GluonFusion}. For both cases Q1 (solid curve) and Q2 (dashed curve), the production rates are significant only for a light flavon, $m_{A_f}^{}\lesssim 200$ GeV, where we take $\kappa_{tt}\sim 0.9$ GeV for Q1, and $\kappa_{tt}\sim 4.5$ GeV for Q2. This is because $\kappa_{tt}$ is suppressed by combination of $U(1)$ charges, but the effect of the bottom quark loop is still sizable for smaller flavon masses. We also show the case of $\kappa_{ff}=m_f$ (dotted curve) for comparison. When the sum of $U(1)$ charges for $Q_3$ and $t_3$ takes non vanishing value, $\kappa_{tt}$ naturally then becomes of order of $m_t$. In this case, the contribution from the top quark loop can be significant, and it would then give the largest cross section.
The flavon can also be generated by the FCNC single top production process $gu\to t A_f$ at tree level, since a proton has a larger distribution of up quarks \cite{Barger:1993th}. The partonic cross section, where helicity and spin are averaged for initial particles but the color index is only summed, is given by
\begin{align}
{\hat \sigma}_{gu\to t A_f} =& \frac{G_F\alpha_S^{}}{\sqrt2{\hat
s}} \left(\left|\kappa_{13}^u\right|^2+\left|\kappa_{31}^u\right|^2\right)
\left(\frac{v}{v_\varphi}\right)^2\nonumber\\
&\times\left\{
2\left[1+2\,x_{t{A_f}}^{}\left(1+x_{t{A_f}}^{}\right)
\ln\left(\frac{1+x_{t{A_f}}^{}+\beta_{t{A_f}}^{}}
{1+x_{t{A_f}}^{}-\beta_{t{A_f}}^{}}\right)\right]
-\beta_{t{A_f}}^{} (3+7\,x_{t{A_f}}^{})\right\},
\end{align}
where $x_{t{A_f}}^{}=(m_t^2-m_{A_f}^2)/{\hat s}$ and
$\beta_{t{A_f}}^{}=\lambda^{1/2}(m_t^2/{\hat s},m_{A_f}^2/{\hat s})$ with $\lambda(x,y)=1+x^2+y^2-2x-2y-2xy$.
In Fig.~\ref{FIG:GluonFusion}, we show the Hadronic production cross sections of the FCNC flavon production for cases Q1 (long-dashed) and Q2 (short dashed) by taking a convolution with CTEQ6M parton distribution \cite{cteq6pdf}.
Because of the large top FCNC coupling, the process $gu\to t A_f$ can be significant, particularly for the case Q2, even if $\kappa_{tt}$ is small.
For case Q2, a wide range of values for $m_{A_f}^{}$ can be accessible at the LHC, while for case Q1 the cross section can be greater than $1$ fb for $m_{A_f}^{}\lesssim 400$ GeV.
Therefore this production mechanism could be important at the LHC for a wide range of flavon masses as can also be seen in Fig.~\ref{FIG:GluonFusion}. We also note that $gg\to tcA_f$ could be substantial since $\kappa_{tc}$ is the largest coupling in the cases Q1 and Q2.

Finally, we study the possible flavon decays. At tree level, the flavon can only decay into fermions but the loop induced decay ${A_f}\to gg$ cannot be neglected in some parameter region, when $\kappa_{tt}$ is order of $m_t$. These leading decay channels are given by,
\begin{align}
&\Gamma_{{A_f}\to f_i{\bar f}_j}^{} = N_C^f
\frac{G_Fm_\varphi}{8\sqrt2\pi}
\left(\left|\kappa_{ij}^f\right|^2+\left|\kappa_{ji}^f\right|^2\right)
\left(\frac{v}{v_\varphi}\right)^2\beta_{ij}\left[1-\frac{(m_i-
 m_j)^2}{m_{A_f}^2}\right],\\
&\Gamma_{{A_f}\to gg}^{} =
\frac{G_F\alpha_S^2m_\varphi^3}{64\sqrt2\pi}
\left(\frac{v}{v_\varphi}\right)^2 \left|N_C\sum_{q}\kappa_{qq}^q
\frac{I(m_q)}{m_q}\right|^2,
\end{align}
where
\begin{align}
&I(m_f) =
4m_f^2C_0(0,0,m_{A_f}^2,m_f^2,m_f^2,m_f^2),
\end{align}
where $\beta_{ij}=\lambda^{1/2}(m_i^2/m_{A_f}^2,m_j^2/m_{A_f}^2)$ and $C_0$ is the scalar loop function whose definition can be found in \cite{Passarino:1978jh}.

\begin{figure}[tb]
\includegraphics[width=10cm]{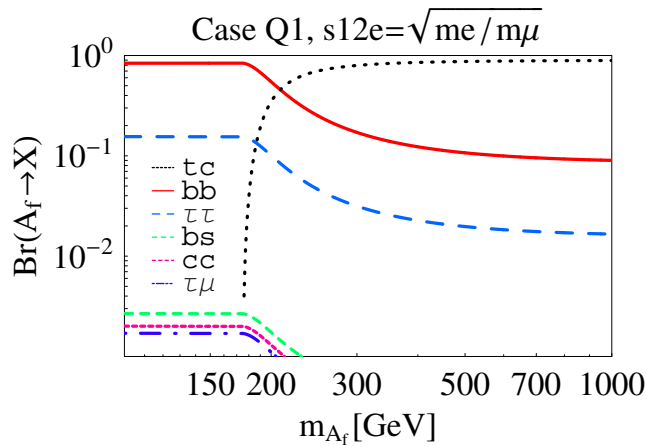}
\caption{Branching ratios for the various flavon decay modes as a function of its mass for the examples detailed in  $\S$ \ref{subs:MIA_1_flavon}.
The decay modes  $tc=\{ t \bar c,\, \bar t c\}$ (dotted), $b\bar b$ (solid), $\tau^+\tau^-$ (long-dashed), $bs$ (dashed), $c\bar c$ (short-dashed), $\tau\mu$ (dot-dashed) are shown. }
\label{FIG:wPhi}
\end{figure}

We present the branching ratio of the flavon in Fig.~\ref{FIG:wPhi} as a function of its mass, note that the total decay width of the flavon simply scales as $1/v_\varphi^2$.
The flavon has a relatively narrow width compared to the SM Higgs boson because it has no tree level gauge
interactions in our setup, and the branching ratio is independent of the vacuum expectation value $v_\varphi$.
The flavon mainly decays into the heaviest fermion-pairs which are kinematically allowed but unlike the Higgs boson its decay modes include flavour changing processes. Hence the fermion-pairs $\{ t \bar c,\, \bar t c\}$, which we call collectively $tc$ pairs, can be the dominant decay products of the flavon for $m_{A_f}>m_t$. For cases Q1 and Q2, it is also a characteristic feature that the decay rate into top-pairs is suppressed by the small $\kappa_{tt}$ values considered. Below the $tc$ threshold, the flavon mostly decays into flavour conserved pairs, i.e., $b\bar b$ and $\tau^+\tau^-$. However, the branching ratios of the LFV decays ${A_f}\to \tau^\pm\mu^\mp$ are not too small and therefore these decay channels could be a useful tool to identify light flavons at the LHC. We also note that detailed studies of LFV Higgs decays at collider can be applied for the flavon LFV decays\cite{lfvhiggslhc,lfvhiggsilc}. We also note that chirality measurements of such FCNC coupling could be important because of $\kappa_{ij}\ne \kappa_{ji} (i\ne j)$, and this can be precisely measured at an electron photon collider by using the electron polarization \cite{lfvhiggsegam}.

%
\section{Conclusions}
We have considered the phenomenology of a flavon $\varphi$ introduced as the scalar that, together with an explicit symmetry breaking mass term, breaks an Abelian non supersymmetric family symmetry giving rise to the hierarchy of the fermion masses and mixing angles. At the scale of the symmetry breaking, $M_F\sim 1$ TeV, this scalar induces flavour changing processes that are controlled by the flavour violating parameters, $\kappa^f$, which have a non-trivial structure in the basis where the Yukawa matrices are diagonal. These parameters are of course closely controlled by the hierarchy in the Yukawa matrices and hence can intimately probe the hierarchy of fermion masses and mixing described by a particular family symmetry in the TeV range.

In this context, we have found that $U(1)_{\text {F}}$ gauged symmetries at the TeV scale are severely constrained by both the cancellation of anomalies and the bounds from LEP given by effective contact interactions mediated by extra $Z^\prime$ bosons. In this context for a group $G_{\text{F}}=U(1)_{\text {F}}$ coupling to all the SM fermions we need to satisfy the bound $m_{\varphi}/\lambda_{\varphi}\sim M_{Z^\prime}/g_F\geq O(100)$ TeV.  These kind of bounds are in contrast to $\sim 1$ TeV \cite{Carena:2004xs} for $U(1)$ models with equal charges for fermions of different families. Even when $\lambda_{\varphi}$ and $g_F$ are taken to be small and there could be boson masses at the TeV scale, most part of the physical processes mediated by the scalar $\varphi$ and the extra gauge boson $Z^\prime$ depend on such bound and hence is not relevant for our study. For $G_{\text{F}}=U(1)_{\text{F}_1}\times U(1)_{\text{F}_2}$, with some of the SM fermions coupling to $U(1)_{\text{F}_1}$ and others to $U(1)_{\text{F}_2}$,  the lightest scalar and gauge boson associated to it needs to satisfy the bound $m_{\varphi}/\lambda_{\varphi}\sim M_{Z^\prime}/g_F\geq 26$ TeV, which is also out of the reach of the LHC.

We introduce then an effective global $G_{\text{F}}=U(1)_{\text {F}}$, broken explicitly by a CP odd mass term (hence avoiding a massless Nambu-Goldstone boson) at the TeV  scale in order to explore the possibility of flavour violating processes, flavon production and decays within the reach of the LHC. These are controlled by a PNGB which is the CP odd part of the scalar $\varphi$, because we have chosen the mass of this, $M$, to be lighter than the vaccum expectation value,  $v_\varphi$, of the whole scalar. The construction of models satisfying this constraint it is left for a future work, here we just have made an account on the theoretical constraints that are relevant and outlined its effective Lagrangian which is constrained to reproduce the appropriate masses and mixing for quark and leptons. For simplicity we have not explored the structure of possible heavy-right handed neutrinos.


Given the current bound on the branching ratio for the flavour violating decay $\mu\to e\gamma$, $B(\mu\to e\gamma)\leq 1.2 \times 10^{-11}$ at the $90$ \% C.L.\cite{exp:muegamma} and its expected improvement \cite{Cattaneo:2009zz}, for some cases we find these models severely constrained if we want to keep values of both $m_{A_f}$ and $v_\phi$ below 1 TeV. Would the flavour structure in the charged lepton sector be controlled entirely by the $U^*_\text{PMNS}$ matrix (L1 case) for $v_\varphi=500$ GeV and $m_{A_f}=200$ GeV then we would be very close to the present limit, as can be seen from Fig.~\ref{FIG:FN-MuEGamma}.
The case L2 is safe above about $m_{A_f}\gtrsim 200$ GeV when $v_\varphi=500$ GeV. A large top-quark mass in the SM is natural, therefore we have just presented examples of $U(1)_F$'s such that the top-quark does not couple at tree level to the flavon. For this reason the flavour violating parameter $\kappa_{tt}$ is very small, $m_t\lambda^2$, and hence the branching ratios of flavour changing decays of top quarks such as $t\to cg,~ c \gamma, c Z$ could by at most $O(10^{-10})$ and therefore out of experimental reach both at the LHC and the ILC.

The branching ratio of $b \to s\gamma$ is largely insensitive to the flavon that we have considered in $\S$ \ref{subs:MIA_1_flavon}, as its contribution to the SM Wilson Coefficient at the decay scale, $C_7(\mu_b)$, for $v_\varphi \geq 150$ GeV  would be of only $\mathcal{O}(10^{-2}\%)$ percent of the SM value, to which has to be added, and therefore cannot alter the SM branching ratio, which at NNLO QCD level is $(3.15\pm 0.23)\times 10^{-4}$ \cite{sm:bsg}. For values of about  $v_\varphi \sim 100$ GeV we can have corrections at the percent level but such values for  $v_\varphi$ are not realistic within our scenario.


For relatively light values of $m_{A_f}\sim 150$ GeV and $v_\varphi=500$ GeV, the decay branching ratios for the processes $t\to c{A_f}$ and $t\to u{A_f}$ it is within the reach of the LHC sensitivity, $\mathcal{O}(10^{-5})$. For the case Q2 introduced in $\S$ \ref{subs:oneflavonquarks}, $\kappa_{tu}$ and $\kappa_{tc}$ respectively correspond to $\lambda^3 m_t$ and $\lambda^2 m_t$.  These values are sufficiently large to find both FCNC top quark decays. While for the case Q1 only the decay $t\to c{A_{f}}$, where $\kappa_{tc}\sim \lambda^2m_t$, is accessible because of the suppressed $\kappa_{tu}\sim \lambda^4m_t$ coupling.

As it happens with the SM Higgs boson, we can expect that the main production channel of the flavon, when $\kappa_{tt}\sim m_t$ is sufficiently large, would be the gluon fusion mechanism at a high energy Hadron collider. Roughly speaking, the production cross section scales as $(\kappa_{tt}/m_t)^2(v/v_\varphi)^2$ (where $v$ is the vev of the Higgs boson) when compared to the SM Higgs production. However, a light flavon can be produced due to the bottom quark loop contribution even if $\kappa_{tt}$ is small.

When $\kappa_{tt}$ is small, $\sim\lambda^2m_t$, the FCNC single top production process $gu\to tA_f$ would be important for relatively heavy flavons, $m_{A_f}\sim 300$ GeV for the case Q1. Again the case Q2 of this production mechanism is very promising for a wide range of flavon masses because of the possibility of a large $\kappa_{tu}$ coupling, which could be order of $m_t\lambda^2$.
This FCNC flavon production would be a complementary process to the gluon fusion mechanism at the LHC.


Finally we have found that below  $m_{A_f}=m_t+m_c$, the main decay modes of flavons are the pair $b\bar b$ and secondly the decay $\tau\bar\tau$. The flavour conserved decay modes are not a particular feature of flavons, but we could nevertheless detect them via these decay modes. In addition, the LFV decay mode $A_f\to \tau\mu$ could be used to identify the flavon, since its branching ratio is of $0.2\%$.

The decay mode $t\bar c$ starts to be the leading one just below $m_{A_f}=m_t+m_c $ GeV  and above it becomes the leading one at more than, for example,  $80\%$ for $m_{A_f}^{}>300$ GeV. This is specific value is for the case Q1 (which has a GST-like mixing in the lepton sector). These decay patterns would hold for the Q2 case, but the only difference would be the opening of the $t u$ decay mode. In any case, flavons can be found in its flavour violating decay if it is sufficiently produced.

\section*{Acknowledgments}

We would like to thank M. Mondrag\'on, L. Covi, A. Ringwald and I. Dorsner for useful discussions. We especially thank J. Kersten for helpful comments and careful reading of the manuscript.

\appendix
\section{Determination of $\mathbf{U(1)_F}$ charges \label{Ap:U1fg_ch}}
\subsection{Global $U(1)_F$ symmetry}

\subsubsection{Quark sector}

Assuming that ${U_L^d}\simeq V_\text{CKM}$, the observed hierarchy of down quark masses and mixing angles is compatible with the following mass matrix \cite{King:2004tx}
\bea
M^d=m_b
\left[
\begin{array}{ccc}
\leq \bep^4 & \bep^3 & \leq \bep^3\\
\leq \bep^3 & \bep^2 & \bep^2\\
\leq \bep & \leq 1   &   1
\end{array}
\right],
\label{eq:M_d_hi_str}
\eea
where $\bep\approx 0.15$. We then see that $U_R^d$ could have the same form of $U_L^d$ and hence $M^d$ could be symmetric. However this is not the only choice, as we can see that $U_R^d=\mathbf{1}$ or have a large mixing in the $(2,3)$ sector. Due to the strong hierarchy of up-type quark mass eigenvalues it is not possible to determine accurately the form of $M^u$, however once $M^d$ is fixed by \eq{eq:M_d_hi_str} with a particular choice of $U_R^d$ then we can restrict the form of $M_u$. If we focus on $U_R^d\simeq U_L^d$ or have only one large mixing in the $(2,3)$ sector, the form of $M^u$ is
\bea
M^u=m_t
\left[
\begin{array}{ccc}
\leq \ep^4 & \ep^3 & \leq \ep^3\\
\leq \ep^3 & \leq \ep^2 & \ep^2\\
\leq  \ep^2 & \leq \ep   &   1
\end{array}
\right],
\label{eq:M_u_hi_str}
\eea
with $\ep\approx 0.05$. A more conservative approach does not necessarily involves assuming $U_R^d\simeq U_L^d$, allowing for the possibility of also having important contributions from the $u$ sector to the CKM matrix, then $Y^d$ and $Y^u$ are constrained to be of the form
\bea
Y^d=a_b
\left[
\begin{array}{ccc}
\leq \lambda^6 & \leq \lambda^5 & \leq \lambda^5\\
\leq \lambda^5 & \lambda^4 & \lambda^4\\
\leq \lambda & \leq \lambda^2   & \lambda^2
\end{array}
\right], \quad
Y^u=a_t
\left[
\begin{array}{ccc}
\leq \lambda^8 & \lambda^6 & \leq \lambda^6\\
\leq \lambda^6 & \leq \lambda^2 & \lambda^2\\
\leq \lambda &  \lambda^2  & 1
\end{array}
\right],
\label{eq:Yf_hier}
\eea
where $a_b$ and $a_t$ are $O(1)$. In our set up for the continuous global $U(1)_F$ the charge of the Higgs boson is taken to be zero and hence the combination $\Fmc^{\oQ}_3+{\Fmc_R^u}_3$ must vanish, other than this and \eq{eq:Yf_hier}, the charges are not restricted by anomaly cancellation conditions like in the gauged $U(1)_F$ case. Requiring a symmetric matrix in the $u$ sector then the matrix of charges acquires the form
\bea
{\mathcal{C}}(Y^u)=\left[
\begin{array}{ccc}
2(s-t) & s &s-t\\
s      & 2t & t\\
s-t    & t  & 0
\end{array}
\right],
\eea
where $(s,t)=(6,2)$ and the charges of the fermions are given by
$\Fmc^{\oQ}_3=-{\Fmc_R^u}_3$, $\Fmc^{\oQ}_2 =t-{\Fmc_R^u}_3$,
$\Fmc^{\oQ}_1=s-t-{\Fmc_R^u}_3$, ${\Fmc_R^u}_2=t+{\Fmc_R^u}_3$,
${\Fmc_R^u}_1=s-t+{\Fmc_R^u}_3$ where ${\Fmc_R^u}_3$ is here a free
parameter. The structure of $Y^d$ does not necessarily needs to be symmetric, even if $Y^u$ it is. Defining $r=\Fmc^{\oQ}_3+ {\Fmc_R^d}_3$ and $t_d=\Fmc^{\oQ}_2+ {\Fmc_R^d}_3$, we present two plausible parameterisations that are in agreement with \eq{eq:Yf_hier}. The first one is
\bea
{\mathcal{C}}(Y^d)=\left[
\begin{array}{ccc}
s-t-{\Fmc_R^u}_3 + {\Fmc_R^d}_1 \quad & t-{\Fmc_R^u}_3 + {\Fmc_R^d}_1 \quad &s-t+r\\
t-{\Fmc_R^u}_3 + {\Fmc_R^d}_1    & 2t+r &  t_d\\
s-t+r    & t_d  & r
\end{array}
\right],
\eea
For $(s,t,r,t_d)=(4,1,3,4)$ we have the solution Q1 presented in Table II. 
 Another possible parameterisation is
\bea
{\mathcal{C}}(Y^d)=\left[
\begin{array}{ccc}
s-t-{\Fmc_R^u}_3 + {\Fmc_R^d}_1 \quad & s-t+r \quad &s-t+r\\
 s-t+r  & t_d & t_d\\
-{\Fmc_R^u}_3 + {\Fmc_R^d}_1    & r  & r
\end{array}
\right],
\eea
With this choice and $(s,t,r,t_d)=(6,2,2,4)$ we have the solution Q2 presented in Table II 
 Note that in either case ${\Fmc_R^u}_3$ remains as a free parameter.
\begin{table}
\label{tbl:U1_global_ch_ex_qks}
\begin{tabular}{|l| l  l l l l l l l l |} \hline
\multicolumn{10}{|c|}{Global $U(1)$ charges for quark ansatzs} \\ \hline
Field      & $Q^\dagger_1$ & $Q^\dagger_2$& $Q^\dagger_3$ & $u_1$ & $u_2$ & $u_3$ & $d_1$   & $d_2$         & $d_3$  \\
 Charge  &
$\Fmc^{\oQ}_1$ &
$\Fmc^{\oQ}_2$ &
 $\Fmc^{\oQ}_3$ &
$\Fmc^{u}_1$ &
$\Fmc^u_2$ &
 $\Fmc^u_3$ &
 $\Fmc^d_1$ &
$\Fmc^d_2$ &
 $\Fmc^d_3$ \\ \hline
Q1 & $4-\Fmc^u_3\quad $ & $2-\Fmc^u_3 \quad $  &  $-\Fmc^u_3 \quad $ &  $4+\Fmc^u_3 \quad $ & $2+\Fmc^u_3 \quad $ & $\Fmc^u_3 \quad $ & $4+\Fmc^u_3 \quad $ & $2+\Fmc^u_3 \quad $ & $2+\Fmc^u_3 \quad $ \\
Q2 & $3-\Fmc^u_3\quad $ & $1-\Fmc^u_3 \quad $  &  $-\Fmc^u_3 \quad $ &  $3+\Fmc^u_3 \quad $ & $1+\Fmc^u_3 \quad $ & $\Fmc^u_3 \quad $ & $6+\Fmc^u_3 \quad $ & $4+\Fmc^u_3 \quad $ & $3+\Fmc^u_3 \quad $ \\
\hline
\end{tabular}
\caption{\small $U(1)_F$ charges for quarks for the global symmetry example.}
\end{table}

\subsubsection{Lepton sector}
In the lepton sector we assume that we have the mass Lagrangian
\bea
-{\mathcal{L}}= Y^e_{ij}\overline{L}_i H {e_R}_j + Y^\nu_{ij}\overline{L}_i H {\nu_R}_j + Z^{\nu}_{ij} {\bar{\nu}^c_{Ri}} \varphi^\prime {\bar{\nu}^c_{Rj}} \ + \ h.c.,
\eea
where the form of $Y^e$ and $Y^\nu$ is given by \eq{eq:mass_lag}. We can assume a direct coupling of the right-handed neutrinos to the flavon $\varphi^\prime$. When this have a vev larger than the TeV scale we can have a see-saw mechanism and so the mass of the left-handed Majorana low energy neutrinos is given by $-m^\nu_{LL}=v^2 Y^\nu M^{-1}_{RR} {Y^\nu}^{T}$, where
\bea
Y^\nu=
\left[
\begin{array}{ccc}
c_{11} \lambda^{{\Fmc_L^\ell}_1} & c_{12} \lambda^{{\Fmc_L^\ell}_1} & c_{13} \lambda^{{\Fmc_L^\ell}_1}\\
c_{21} \lambda^{{\Fmc_L^\ell}_2} & c_{22} \lambda^{{\Fmc_L^\ell}_2} & c_{23} \lambda^{{\Fmc_L^\ell}_2}\\
c_{31} \lambda^{{\Fmc_L^\ell}_3} & c_{32} \lambda^{{\Fmc_L^\ell}_3} & c_{33} \lambda^{{\Fmc_L^\ell}_3}
\end{array}
\right], \quad
(M_{RR})_{ij}=v_{\varphi} Z^\nu_{ij},
\eea
note however the examples considered in $\S$ \ref{subsc:leptsect} are non sensitive to this scale and we can just account for the charges of lepton sector. In order to reproduce the correct hierarchy of charged lepton masses the charges $\Fmc^{\oL}_i$ need to satisfy
\bea
|{\Fmc^\ell}_1| > |{\Fmc^\ell}_2| >| {\Fmc^\ell}_3|,
\eea
for $\ell=\oL, e$. This condition alone turns out to be quite restrictive. Then the charges of $Y^e$ can be parameterised as
\bea
 {\mathcal{C}}(Y^e)=\left[
\begin{array}{ccc}
p^e_{11}                  &  p^e_{11} + \Fmc^e_2- \Fmc^e_1    & p^e_{11} + \Fmc^e_3- \Fmc^e_1  \\
t_e + \Fmc^e_1- \Fmc^e_3  &  t_e + \Fmc^e_2- \Fmc^e_3         & t_e\\
r + \Fmc^e_1- \Fmc^e_3    &  r + \Fmc^e_e- \Fmc^e_3           & r
\end{array}
\right].
\label{eq:chYepar}
\eea
For the case L1 where $U^e_R=U^*_{PMNS}$ we have the following solution
\bea
 {\mathcal{C}}(Y^e)=\left[
\begin{array}{ccc}
p^e_{11}                  &  p^e_{11} + \Fmc^e_3- \Fmc^e_1    & p^e_{11} + \Fmc^e_3- \Fmc^e_1  \\
t_e + \Fmc^e_1- \Fmc^e_3  &  t_e          & t_e\\
r + \Fmc^e_1- \Fmc^e_3    &  r           & r
\end{array}
\right],
\eea
that is $\Fmc^e_2=\Fmc^e_3$. Note that once $p^e_{11}$ is specified $\Fmc^{\oL}_1$ is uniquely determined but all other charges work for an arbitrary value of $\Fmc^e_3$, just as in the quark case.

The solution L2, for which $U^e_R=\mathbf{1}$  preserves the general form of \eq{eq:chYepar} since $\Fmc^e_e\neq \Fmc^e_3$ and in this case once $p^e_{11}$ is specified $\Fmc^{\oL}_3$ is uniquely determined but the rest of the charges are defined only up to $\Fmc^e_3$. The charges for solutions L1 and L2 are presented in Table III of this appendix. 
\begin{table}
\label{tbl:U1_global_ch_ex_lept}
\begin{tabular}{|l| l  l l  l l l |} \hline
\multicolumn{7}{|c|}{Global $U(1)$ charges for lepton ansatzs} \\ \hline
Field    &  $L^\dagger_1$ & $L^\dagger_2$ & $L^\dagger_3$  & $e_1$  & $e_2$         & $e_3$ \\
 Charge  &
$\Fmc^{\oL}_1$ &
 $\Fmc^{\oL}_2$ &
 $\Fmc^{\oL}_3$ &
 $\Fmc^e_1$ &
 $\Fmc^e_2$ &
$\Fmc^e_3$\\ \hline
L1 & $6+ {\Fmc^e_R}_3\quad$                & $4-{\Fmc^e_R}_3\quad $ & $4-{\Fmc^e_R}_3\quad $   & $2+{\Fmc^e_R}_3\quad $ & ${\Fmc^e_R}_3\quad $ & ${\Fmc^e_R}_3\quad $ \\
L2 & $5-{\Fmc^e_R}_3\quad $ & $4-{\Fmc^e_R}_3 \quad $ & $4-{\Fmc^e_R}_3\quad $     & $5+{\Fmc^e_R}_3$ & $2+{\Fmc^e_R}_3\quad $ & ${\Fmc^e_R}_3$
\\ \hline
\end{tabular}
\caption{\small $U(1)_F$ charges for leptons for the global symmetry example.}
\end{table}
%
%
%
\subsection{Gauged $U(1)_F$ symmetry}
%

In $\S$ \ref{sbsc:gaguedmod} we have detailed some possible solutions for charges which satisfy the cancellation of anomalies. In Table IV of this appendix we present specific values of these charges and we leave for a future work a detailed analysis of this kind of solutions.
\begin{table}[ht]
\label{tbl:charges_gauged_ex}
\begin{tabular}{|l| l  l l l l l l l l l l l l l l l l l|} \hline
\multicolumn{19}{|c|}{Gauged $U(1)$ charges} \\ \hline
Field      & $Q^\dagger_1$ & $Q^\dagger_2$& $Q^\dagger_3$ & $u_1$ & $u_2$ & $u_3$
 & $d_1$   & $d_2$         & $d_3$         & $L^\dagger_1$ &
$L^\dagger_2$ & $L^\dagger_3$ & $e_1$         & $e_2$         & $e_3$ & $H$ & $\varphi_1$ & $\varphi_2$ \\
 Charge  &
$\Fmc^{\oQ}_1$ &
$\Fmc^{\oQ}_2$ &
 $\Fmc^{\oQ}_3$ &
$\Fmc^{u}_1$ &
$\Fmc^u_2$ &
 $\Fmc^u_3$ &
 $\Fmc^d_1$ &
$\Fmc^d_2$ &
 $\Fmc^d_3$ &
$\Fmc^{\oL}_1$ &
 $\Fmc^{\oL}_2$ &
 $\Fmc^{\oL}_3$ &
 $\Fmc^e_1$ &
 $\Fmc^e_2$ &
$\Fmc^e_3$ &
${\mathcal{Q}}_H$ &
${\mathcal{Q}}_{F_1}$ &
${\mathcal{Q}}_{F_2}$
\\ \hline
$U(1)_{F_1}$ & 4 & 2 & 0 & 4 &2 & 0 & -24 ~ & 3 & 3 & $-\frac{512}{21}$ ~ & $\frac{155}{42}$ ~ & $\frac{113}{42}$ ~ & $\frac{134}{21}$ ~& $\frac{13}{42}$ ~ &  $-\frac{29}{42} ~$ & 0 & -1 & 0 \\
$U(1)_{F_2}$ & 0 & 0 & 0 & $-\frac{20}{3}$ & $-\frac{1}{3}$ & $-\frac{2}{3}$ & 7 & 0 & 0 & 4 & $-\frac{3}{2}$ & $-\frac{5}{2}$ & 3 &3 & 1 & 0 & 0 & -1 \\ \hline
$Z_2$ & 0 & 0 & 0 & 0 & 0 & 0 &1 &1 &1 & 0 & 0 &1 &1 &0 &1 & 0 & 1 &1\\ \hline
\end{tabular}
\caption{Charges of the SM fermions and the flavons under the $U(1)_{F_1}$, $U(1)_{F_2}$ and the $Z_2$ groups.}
\end{table}

\section{Determination of flavour violating parameters \label{ap:flvpars}}

In Section \S \ref{subs:MIA_1_flavon}-A we considered the flavour violating parameters $\kappa^f_{ij}$ when the mass matrices are symmetric and hence diagonalised by a unitary matrix and its transpose, consequently the flavour violating parameters are given by \eq{eq:fv_ks_msym}. Since $\kappa^\fsm_{lk}$ is then also a symmetric matrix we have
\bea
\kappa_{lk} && = v \left[\sum_i U_{li} p_{ii}(U^\dagger Y_{\rm{diag}}U^*)_{1i} U_{ki}
+  p_{12}(U^\dagger Y_{\rm{diag}}U^*)_{12} \left(U_{l2}U_{k1}+U_{l1}U_{k2} \right)\right.\nn\\
&&\left. \ +  p_{13}(U^\dagger Y_{\rm{diag}}U^*)_{13} \left(U_{l2}U_{k3}+U_{l3}U_{k1} \right)
   \ +  p_{23}(U^\dagger Y_{\rm{diag}}U^*)_{23} \left(U_{l2}U_{k3}+U_{l3}U_{k2} \right) \right]
\eea
The elements $v(U^\dagger Y_{\rm{diag}}U^*)_{ij}$ can be approximated by
\bea
v(U^\dagger Y_{\rm{diag}}U^*)_{12} &\simeq& U^{\dagger}_{12}m_2 + U^\dagger_{13}U_{32}m_3\nn\\
v(U^\dagger Y_{\rm{diag}}U^*)_{13} &\simeq& U^{\dagger}_{12} U^*_{23} m_2 + U^\dagger_{13}m_3\nn\\
v(U^\dagger Y_{\rm{diag}}U^*)_{23} &\simeq& U^*_{23}m_2 + U^\dagger_{13}U^*_{32}m_3\nn\\
v(U^\dagger Y_{\rm{diag}}U^*)_{11} &\simeq& m_1 + (U^\dagger_{12})^2 m_2 + (U^\dagger_{13})^2 m_3\nn\\
v(U^\dagger Y_{\rm{diag}}U^*)_{22} &\simeq& m_2 + (U^\dagger_{23})^2 m_3\nn\\
v(U^\dagger Y_{\rm{diag}}U^*)_{33} &\simeq& m_3.
\eea
%


\end{document}